\documentclass{sig-alternate}
\usepackage[pagebackref=true,breaklinks=true,colorlinks,bookmarks=false]{hyperref}
\usepackage{balance}  
\usepackage{graphics} 
\usepackage{times}    
\usepackage{url}      
\usepackage[percent]{overpic}
\usepackage{amsmath}

\usepackage{enumitem}
\setlist{nolistsep}

\usepackage{array}

\DeclareMathOperator*{\argmax}{arg\,max}

\begin{document}\sloppy
%
\conferenceinfo{WOODSTOCK}{'97 El Paso, Texas USA}

\title{When relevance is not Enough: Promoting Visual Attractiveness for Fashion E-commerce}

\numberofauthors{5} 
\author{
Wei Di, Anurag Bhardwaj, Vignesh Jagadeesh, Robinson Piramuthu, Elizabeth Churchill\\
       \affaddr{eBay Research Labs}\\
       \affaddr{2145 Hamilton Ave., San Jose, CA, USA}\\
       \email{{wedi, anbhardwaj, vjagadeesh, rpiramuthu, echurchill}@ebay.com}       
}

\date{8 Aug. 2013}

\maketitle
\begin{abstract}
\label{sec:abs}

Fashion, and especially apparel, is the fastest-growing category in online shopping. 
As consumers requires sensory experience especially for apparel goods for which their appearance matters most, images play a key role not only in conveying crucial information that is hard to express in text, but also in affecting consumer's attitude and emotion towards the product. However, research related to e-commerce product image has mostly focused on quality at perceptual level, but not the quality of content, and the way of presenting. 
This study aims to address the effectiveness of types of image in showcasing fashion apparel in terms of its attractiveness, i.e. the ability to draw consumer's attention, interest, and in return their engagement. We apply advanced vision technique to quantize attractiveness using three common display types in fashion filed, i.e. human model, mannequin, and flat. 
We perform two-stage study by starting with large scale behavior data from real online market, then moving to well designed user experiment to further deepen our understandings on consumer's reasoning logic behind the action. 
We propose a Fisher noncentral hypergeometric distribution based user choice model to quantitatively evaluate user's preference.
Further, we investigate the potentials to leverage visual impact for a better search that caters to user's preference. A visual attractiveness based re-ranking model that incorporates both presentation efficacy and user preference is proposed. We show quantitative improvement by promoting visual attractiveness into search on top of relevance.


\end{abstract}

\category{H.2.8}{Database Applications}[data mining, image databases]



\section{Introduction}
\label{sec:intro}


Online marketplaces have grown at scale along with the internet providing opportunities for local, cross-border, and global commerce. Consumers enjoy the convenience and low prices offered by online shopping.
Study suggests that an e-commerce website is a type of ``decision support system'' that supports the stages of the purchasing decision~\cite{miles2000framework}. Multiple factors affect user`s choice in browsing and purchasing. The key factors include trust, logic and emotion, whereas emotion refers to the ability to link the customer to the product and services, or cause the realization at certain level of senses such as feeling and wanting. Reports also show that shopping behavior is significantly influenced by consumer`s attitude, social influence, trust and perceived risk, etc.~\cite{moshrefjavadi2012analysis,chen2009online}.

With no physical items to inspect, consumer's decisions rest purely on the the descriptions and pictures provided~\cite{haywood2006online}. Therefore, effective communication between buyer and seller is very important.
However, text information such as title and listing description can only provide information within the scope of language. Fortunately, product image is shown to provide a unique yet profound channel, to convey visual information to buyer, for which text description may not be capable of. 
In addition to carrying important visual information, image is also recognized as a powerful way in persuasive communication and as a crucial determinant of memory and attitudes. It is also thought to be able to easily grab people`s attention and effectively affect their emotion as compared to its verbal counterpart in the process of persuasion. 
Studies regarding those advantages show that the inclusion of image helps to reduce the perceived risk for e-shoppers~\cite{park2005line}. 
Yet, simply including visual images does not necessarily bring success and ensure the quality of communication. There is one under-researched aspect, which is the effectiveness of product image presentation in the sense of providing shopping enjoyment to influence consumer's attitudes toward product~\cite{jiang2007research}.
Given the huge variation of how similar or same content/product can be presented in different ways, our essential question is what is the best way to present a product using image, such that it is most effective in engaging user, arousing and forming favorable emotion and attitude towards the product, and hence enable pleasing shopping experiences that could lift purchase intentions. To answer such question, it is important to examine how people interpret and evaluate visual information, and how user responses to make the best possible decision. 

These questions are particularly important for apparel category for which sensory evaluations are crucial for making purchase decision~\cite{yoo2012online,bhardwaj2013palette, di2013style}. 
Reports have shown that apparel is becoming the fastest-growing segments in e-commerce, and is expected to become the second biggest segment by revenue overall~\cite{emarket2012}. 
Unlike other categories, consumers often require sensory evaluations through mental imagery for experience goods like apparel. E-shoppers like to get a sense of how the clothing will look like when being wear, and even how the fabrics and texture feels like. Therefore, it is important to gain better understanding of consumer's reaction toward apparel goods in their online image format.

So far, many e-commerce platform have developed various tools to improve product presentation, including zooming, panning, multiple-views, etc. Despite all the new inventions of visualization tools, the bottleneck remains and customer still feel higher risk for online apparel purchase as compared to off-line local store purchase. 

Beside the industry efforts, this issue also draws attention to academic field. Some researchers have been looking at the online visualization tools~\cite{yoo2012online}. Others focus on visual and perceptual level, assessing quality, clarity, visual salience and perceptual attention\cite{moore2011computer}. It has also been found that the more appealing and interesting the product display is, the higher the purchase intentions~\cite{song2012does}. However, despite the great variations in clothing presentation, only few researchers have looked at content quality from product presentation perspective, and its connection with shopping attitude and emotions. So far, related work are still extremely limited. Most of them are based on small user group experiments and are often not targeted for fashion area.

On the other end of spectrum, we would also like to argue that unlike general search engine, where item relevance is the essential evaluation metric, for a successful e-commerce website, only relevance is not enough. User engagement, enjoyable and inspiring experience are also important. 
The goal is not only to find what consumers ``\textbf{want}'', but also to inspire them for what they ``\textbf{like}'', 
In literature, some researchers have proposed exploiting user behavior and feedback as an evaluation signal~\cite{joachims2007search,radlinski2005query,agichtein2006improving}. While behavioral evaluation can be very effective in certain circumstances, given the complexity and dynamics of behavioral data, there is still a lack of precise understanding of what are the driven factors. Therefore, we expect our work to provide useful insights on how user preferences affect their behavior from a vision perceptive.   


\begin{figure}[t]
\centering
\epsfig{file=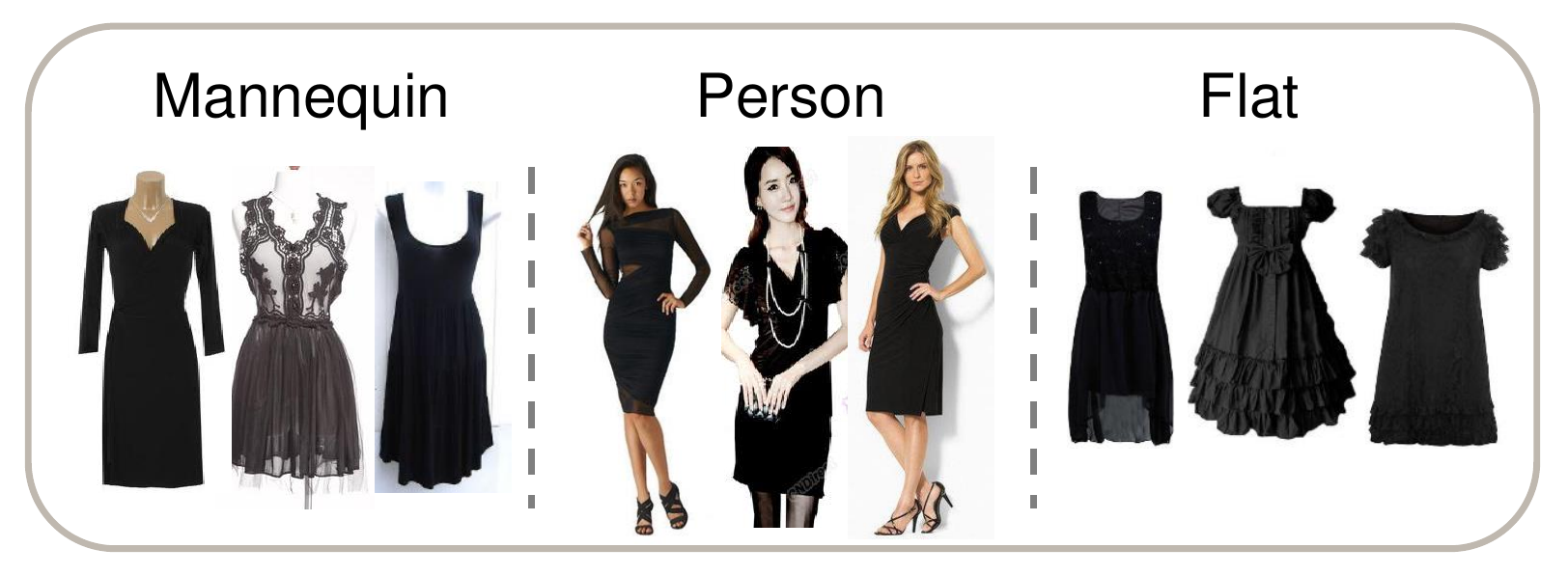,width=0.88\linewidth} \vskip -9pt
\caption{Three display types of fashion clothing. Our question is whether the same article of clothing, e.g. a little black dress, attract user differently depending on how it is represented?}
\label{fig:pmf_illustration}
\end{figure}

Differ from previous research, our work is motivated by the fact that humans are naturally drawn to images of people. As shown in Figure~\ref{fig:pmf_illustration}, overall, we want to investigate situations in which product visual presentation influences user's participating, engaging and purchase behavior, particularly for online fashion clothing commerce. Specifically, the research questions we ask are:
 
\begin{itemize}
\item Does visual presentation influence consumer behavior across different stages of online shopping?
\item How influential such impact is? How much such preference affects user choice during online shopping? 
\item What is the most effective fashion clothing presentation that could draw user's attention, raise their interest and build the connection between product and consumer, in other words, attract user effectively?
\item How we quantitatively measure user preference on the way that clothing is presented?
\item How to utilize the feedback from user on product presentation to improve shopping experience by providing not only what consumer want but also what they like?
\end{itemize}

To answer these questions, contrary to previous work which often studies the quality or perceptual property of the image (salience, clarity, etc), we investigate the ``attractiveness'' aspect of the presentation. The first step is to quantize attractiveness, meaning building a mapping function between the visual presentation (in form of image) with its ability of arousing favorable feelings from user, such as attention and interestingness. 
While attractiveness can be measured via different dimensions, We hypothesis that different presentation types result in different attractiveness levels. Thus, we utilize three common product presentation types that are often seen in online fashion market, namely, {\fontsize{8.0}{9.8}\selectfont \textsf{Person}} (use of human model, abbrev. as P),
{\fontsize{8.0}{9.8}\selectfont \textsf{Mannequin}} (M), and {\fontsize{8.0}{9.8}\selectfont \textsf{Flat}} (F), to represent attractiveness with a three-point scale in a more discernible manner. Through this bridge, we hope to build the connections between attractiveness and user preference and their behavior.
Our goal is to discover and quantitatively measure the visual presentation influence in an online decision making process. Also, we aim to identify the most effective product presentation in terms of attracting user`s interest or attention by utilizing advanced computer vision models built on top of these three types.

Our contributions are as follows:
\begin{itemize}
\item We conduct large-scale analysis over millions of search sessions and items from a worldwide online marketplace. We leverage user behavior and transaction data (click, watch, and purchase) to study the influence and effectiveness of fashion clothing product presentation, in terms of attracting user`s interest or attention. 

\item A survey based user study is presented to deepen the understandings of motives of user choice by separating vision information with other meta information associated with the product. By carefully designing the experiment, we gain clearer insights about the reasoning logic and other important factors that affect consumer's choice, the gender discrepancy, and impacts from price and brand with respect to purely visual information. 

\item We propose a Fisher noncentral hypergeometric distribution based user choice model, which can quantitatively measure the preference level that learned from user click data. 

\item We propose a visual attractiveness based re-ranking model. We show quantitatively that by incorporating attractiveness element into search engine, in the form of re-ranking, we can promoting better user engagement on top of relevance. 

\end{itemize}

This is the first study focuses on factors that people find attractive for online apparel goods. We believe this study can be highly useful for a number of applications. In addition, our findings may also have implications for content designers, sellers, advertisers, and so on, who want to attract attentions of clothing/apparel. 
Although this study only focus on fashion e-commerce, we believe similar concept and framework can be generalized to other domains. It is recommended to consider individual characteristics from each domain in terms of what are the preferred visual effect.

This paper is organized as follows:  In section~\ref{sec:related-work}, we give a brief review on related work in literature. Section~\ref{sec:vision} begins by first describing data collection and the approach used to quantize attractiveness by categorizing image content using a proposed {\fontsize{8.0}{9.8}\selectfont \textsf{PMF}} image classification model. We then present the main insights drawn from the real online marketplace in section~\ref{sec:data-anatomy}, followed by survey based user study in section~\ref{sec:user-study}. 
Section~\ref{sec:weighing} proposes a user choice model to quantitatively evaluate user's preference by modeling click data using Fisher noncentral hypergeometric distribution. The visual attractiveness based reranking model and experiment results are given in section~\ref{sec:search-reranking}. The last section~\ref{sec:conclusion} concludes the paper and discusses potential applications.

\section{Related Work}
\label{sec:related-work}

Extensive studies of online marketplaces have been done on consumer behavior, selling strategy, trust and other related issues~\cite{chen2009online}. 

Previous research has shown the importance of use of pictures for buyers on one of the largest e-commerce market - eBay~\cite{bajari2000winner, bland2005determinants,lohse1998electronic,gilkeson2003determinants,trocchia2000phenomenological}.  
Study in~\cite{bland2007risk} indicates that product picture is one of the most influential risk-reducing factors.
The author found in their particular case that either a real picture of the product actually being sold or a stock picture is a risk-reducing factor that will improve the outcomes of the auction. The inclusion of the real picture is proved to be effective in increasing auction success, effectiveness, and the value of the final bid. While a stock picture also significantly increases the final bid, the probability of auction success is not enhanced.  
Study in~\cite{lewis2009asymmetric} looks at number of images embedded in item descriptions in eBay motors and draws the conclusion that more images help to boost the selling, especially for old cars.
The author also found strong correlation of photo and price for non-dealers, which is possibly because buyers cannot rely on reputation as an alternate source of information about quality.
Moreover, evidence has been presented that clear and detailed pictures of the products also help to reduce perceived risk associated with online purchasing~\cite{helander2000theories,wolfinbarger2001shopping,koehn2003nature}. 
A study on a smaller user group reveals the importance of better quality images as a number of shoppers expressed their preference to see both higher quality photographs and more images in item descriptions~\cite{keynote2005}. 
		
Recent study showed that certain image features can also help to improve click through rate in product search engine~\cite{goswami2011study, chung2012impact}. The authors conducted experiments to show that including image features in a machine learned click based ranking model improves the NDCG (normalized discounted cumulative gain) of the search results.
The work in ~\cite{bland2005determinants} studies the ability to attract customers and likelihood of transaction by including a binary variable indicating whether an image is associated with the listing. They found positive evidence for inexpensive products that providing image can increase number of bidding. 

All the above studies address the importance of image in online shopping network from different aspects. However, most of them focus on the impact of inclusion of image, or low level image quality, but not the content of image, and the way of presenting. Only handful of studies have been trying to look the problem by understanding user preference from psychological point of view. 

Study in advertising often supports using of pictures of women~\cite{karlin2005s}, but these experiments are not for online market. 
Kim \textit{et al}. studied the effect of use of a model, color swapping and enlargement on emotions, such as pleasure and arousal, with respect to perceived risk~\cite{kim2010atmosphere}. They found positive relationship between pleasure and perceived amount of information, and negative relationships between perceived amount of information and product quality and online transaction risk and consequential risk.
Another study~\cite{yoo2012online} examined the effects of product coordination and a model`s face on consumer responses in terms of affecting states, perceived amount of information and purchase intention. Out of expectation, they found that consumers perceived more information when no model's face was present with the product than when an attractive model's face and body were shown together. 
While these work are limited to small user group with subjective ratings, which lack of completeness and representativeness of the vast online shoppers in a real marketplace, results are still very exciting and pave the way for future study on product presentation and consumer emotions.  

Contrary to previous study, we aim to propose a general model to evaluate the effectiveness of product images, specifically on fashion clothing images regarding its attractiveness, i.e. the ability to draw consumer's attention, interest, and in return their engagement. In our study, we leverage advanced vision technique to closely look into fashion filed and quantize attractiveness using three different display types that are commonly seen in fashion filed. Not limited to small user group study, we start with conducting large scale data analysis on real online transaction and behavior data, and then design user experiment to further deepen and refine our understandings of user's reasoning logic behind the action. 
We propose a PMF user choice model to quantitatively evaluate user's preference by modeling click data using Fisher noncentral hypergeometric distribution. A new visual attractiveness based re-ranking schema is proposed. The goal is to show the potentials of incorporating visual attractiveness and user preference factors to promote user engagement on top of relevance.

\section{Categorizing Attractiveness}
\label{sec:vision}

\subsection{Data Collection}
We collect user behavior data from one popular e-commerce platform for two periods in 2012.
The site enables user to search for product using text query. 
In each search session, user inputs a query looking for certain product and the search engine returns multiple items ranked by their relevance.  
Since search is a very personalized task with huge variations in terms of search intention and product attributes, in order to focus only on apparel goods and limit images to be of similar content, i.e. the same category of product, we only collect sessions with query containing the keyword ``\textit{black dress}''. Also, we only looked at highly relevant items displayed on the 1st search result page, and which items were clicked. 
By this way, collected images are most likely with the same content - black dress, but only differ in the way of presentation. For example, common ways of displaying an dress is to either use of human model, mannequin or just flat. This helps us focus on the core problem by removing unrelated factors. In total, we collected $29$k search sessions with $429$k images.

\subsection{Representing Attractiveness through Image Classification}

\begin{figure}
\centering 
\epsfig{file=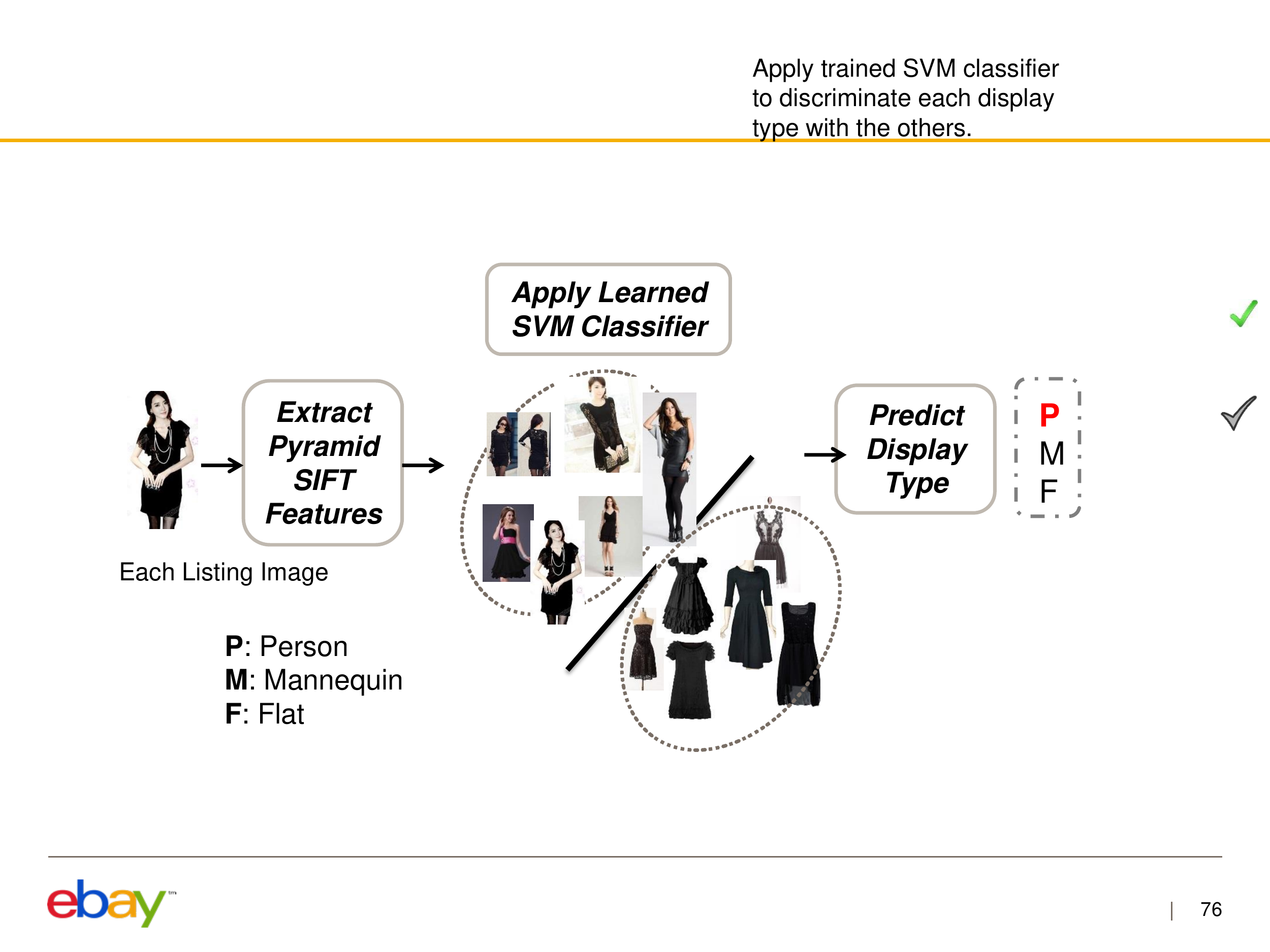,width=0.95\linewidth}  \vskip -9pt
\caption{Illustration of using learned SVM classifier to predict the level of attractiveness categorized by distinct common display types ({\small PMF}) for listing images.} 
\label{fig:svm_classify_illustration} 
\end{figure}

To understand user preference, our first step is to understand the image content and how it relates to attractiveness. 
As aforementioned, modeling attractiveness is a nontrivial task. Here we propose to take advantages of three common clothing presentation types: {\fontsize{8.0}{9.8}\selectfont \textsf{Person}} ({\small P}),
{\fontsize{8.0}{9.8}\selectfont \textsf{Mannequin}} ({\small M}),
or {\fontsize{8.0}{9.8}\selectfont \textsf{Flat}} ({\small F}) in online marketplace to quantize attractiveness in a more discriminative manner. By directly mapping attractiveness to existing display types can also help to identify the most effective way of presentation and in return help content designers in a more straightforward manner. 
To learn such mapping function, we reformat the problem to be an image classification problem with three distinct categories ({\small PMF}). 
From the collected dataset, we randomly sampled $2.4$k images, and obtained image-level annotations using Amazon Mechanical Turk~\footnote{\url{http://www.mturk.com}}. For each image, workers were asked to select one of the categories ({\small PMF}).
Incorrect labels were further manually cleaned. This in total produces $2392$ annotated images (\textsf{\small F}:$881$, \textsf{\small M}:$741$, and \textsf{\small P}:$770$) for building the PMF-attractiveness classification model.

For this multi-class classification problem, we use one-vs-all schema. For each class, a subset of $450$ images are selected as positive examples for training, and all the left images are used as testing set for evaluation purpose. We first extract SIFT features ({\small PHOW}) for each image~\cite{vedaldi08vlfeat}.  
SIFT features are computed densely at three scales on a regular grid and quantized using a bag-of-words model (BoW) with 1000-word vocabulary~\cite{csurka2004visual}. The vocabulary is built by applying \textit{k}-means clustering on all {\small PHOW} features extracted from a set of images with 30 images randomly selected from the training images of each class. Suppose $\mathbf{x}$ is the BoW feature vector extracted from a given image $I$, each of the PMF-classifiers for scoring attractiveness level is:
\begin{equation} \label{eq:Fc}
\mathcal{S}_{c\in{p,m,f}}(\mathbf{x}) = \sum\limits_{i=1}^{N_c} \alpha_{i}^c K(\mathbf{x},\mathbf{x}_{i}^c) + b
\end{equation}
\noindent where $K$ is the kernel function and $N_c$ is the number of support vectors for class $c$. Here we chose chi-square kernel as they perform well for BoW features. $\mathcal{S}_{c\in{m,p,f}}(\mathbf{x})$ is the decision value, whereas a larger positive value implies high confidence of belonging to positive class, i.e. to a certain level of attractiveness. We can obtain the label of level by:
\begin{equation}
\mathcal{L}(\mathbf{x}) = \argmax\limits_{c\in{p,m,f}}{ \mathcal{S}_{c}(\mathbf{x})} 
\end{equation}

Since training a non-linear classifier can be computationally expensive, we therefore firstly map the original features into kernel space using approximate kernel mapping~\cite{vedaldi2012efficient}, which is able to proximate the kernel dot product by using only limited kernel dimension. The kernel approximation function $\hat{\psi}$ is learned through all training features, such that $K(\mathbf{x},\mathbf{x}_i) = \langle \hat{\psi}(\mathbf{x}), \hat{\psi}(\mathbf{x}_i) \rangle$. Denote $\hat{\psi}(\mathbf{x}) $ the mapped feature in the kernel space, the classifier can be transformed to a linear classifier in the approximated feature space:
\begin{equation}
S_{c\subset{p,m,f}}(\mathbf{x}) =  \langle \mathbf{\varpi}, \hat{\psi}(\mathbf{x})  \rangle 
\end{equation}
\noindent where weights of the linear hyperplane $\mathbf{\varpi}=\sum\limits_{i=1}^{N_c} \alpha_i \hat{\psi}(x)$ is the linear combination of all support vectors in the approximated kernel space. 

The final SVM classifier achieves an accuracy of $81.89\%$ on testing dataset. We then use this model as the final predictor to categorize the attractiveness level (three levels: P, M, F) for all $429$k listing images in the collected dataset. Figure~\ref{fig:svm_classify_illustration} shows the pipeline of our system.

\section{Story of Real-world Market}
\label{sec:data-anatomy}

\begin{figure}[t]
\centering 
\epsfig{file=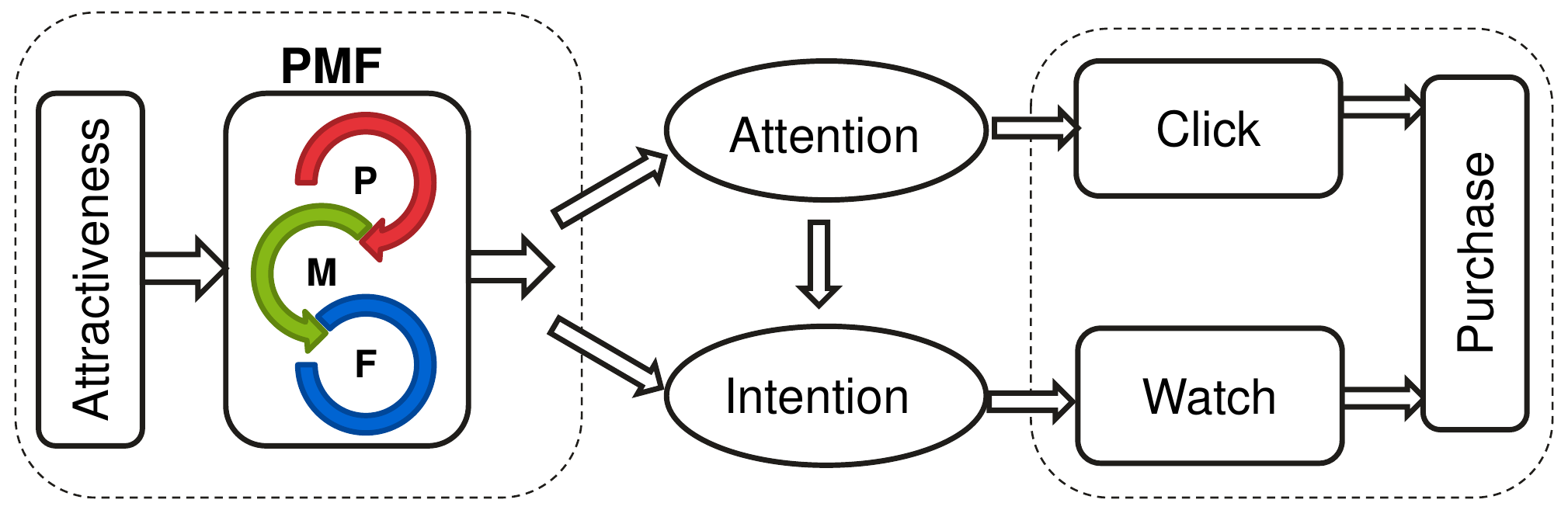,width=0.95\linewidth} \vskip -9pt
\caption{Attractiveness is quantized into three levels, with each corresponds to one type of display mode. Three display types: Person (P), Mannequin (M), Flat (F), differ in the level of attracting consumer (attention \& intention), thus impact user's behavior (click, watch, \& purchase).}   
\label{fig:SOR_model}
\end{figure}

Common to any successful transaction, attract potential consumer`s attention to the product is crucial. User interest can be shown at different stages during the process of online shopping, e.g. browsing, click action and purchase. As shown in the diagram~\ref{fig:SOR_model}, we are interested in using {\fontsize{8.0}{9.8}\selectfont \textsf{PMF}} attractiveness model to understand and quantify user's responses at three stages during the online purchasing circle: a) ``\textit{Click}'' at the search result page, where multiple relevant items are displayed according the search query; b) ``\textit{Watch}'' action at the view item page, where shoppers are in the process of evaluating the item in great detail and make decision to either put it on hold (by watching) or continue to browse or purchase; c) ``\textit{Purchase}'' - user`s final decision on the product.

\subsection{The First Glance: Attention} 
\label{subsec:fist-glance} 

\begin{table}[h] \small
\centering 
\begin{tabular}{c|c|c|c}
\textbf{Type} & \textbf{Displayed Items} & \textbf{Clicked Items} & \textbf{Unclicked Items} \\
\hline 
Flat		& 40.87 \% 	& 39.21  \%	 & 40.99\%  \\
Mannequin		& 34.49 \%	& 33.26  \%	 & 34.57\%	\\
Person			& 24.65 \%	& \textbf{27.53\%}	& 24.44\%	\\
\end{tabular}
\caption{Distribution shift for displayed, clicked, \& unclicked items. For clicked items, proportion of P-type increases while M and F-type decreases indicating users favor P-type over M or F-type.}  
\label{tab:dist_shift_showed_clicked_unclicked}
\end{table}

Given multiple relevant items displayed on the search result page, user's click response can be affected by various factors shown on the page, including relevance, prices, images and their displaying format, seller information, etc.
By categorizing image content into {\fontsize{8.0}{9.8}\selectfont \textsf{PMF}} types, each representing different levels of attractiveness, Table~\ref{tab:dist_shift_showed_clicked_unclicked} shows significant distribution shift from the original displayed search result to what were clicked by the users. Ratio of {\fontsize{8.0}{9.8}\selectfont \textsf{Person}}-type (P-type afterwards) is only $24.65\%$ for retrieved items, but increases to $27.53\%$ for clicked items. Proportions decrease for both {\fontsize{8.0}{9.8}\selectfont \textsf{Mannequin}} and {\fontsize{8.0}{9.8}\selectfont \textsf{Flat}} types for clicked items. Clearly, this shows that users tend to click more on P-type, which indicates more attention is drawn for P-type as compared with M-type and F-type.

\begin{figure}[h]
\centering \vskip -9pt
\begin{tabular}{c}
\epsfig{file=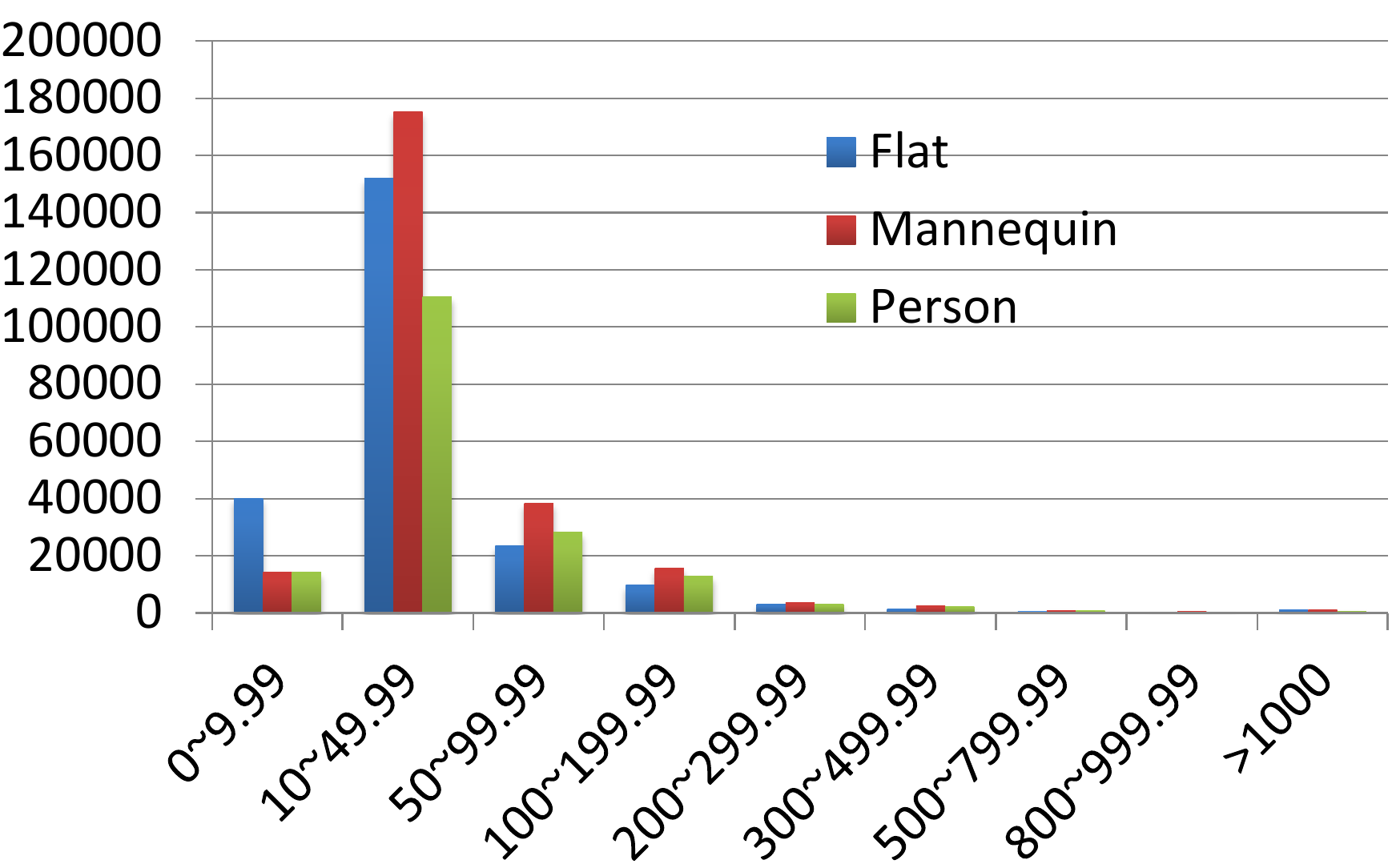,width=0.8\linewidth}  \\
{\scriptsize (a) Distribution of unclicked items over various price segments}  \\
\epsfig{file=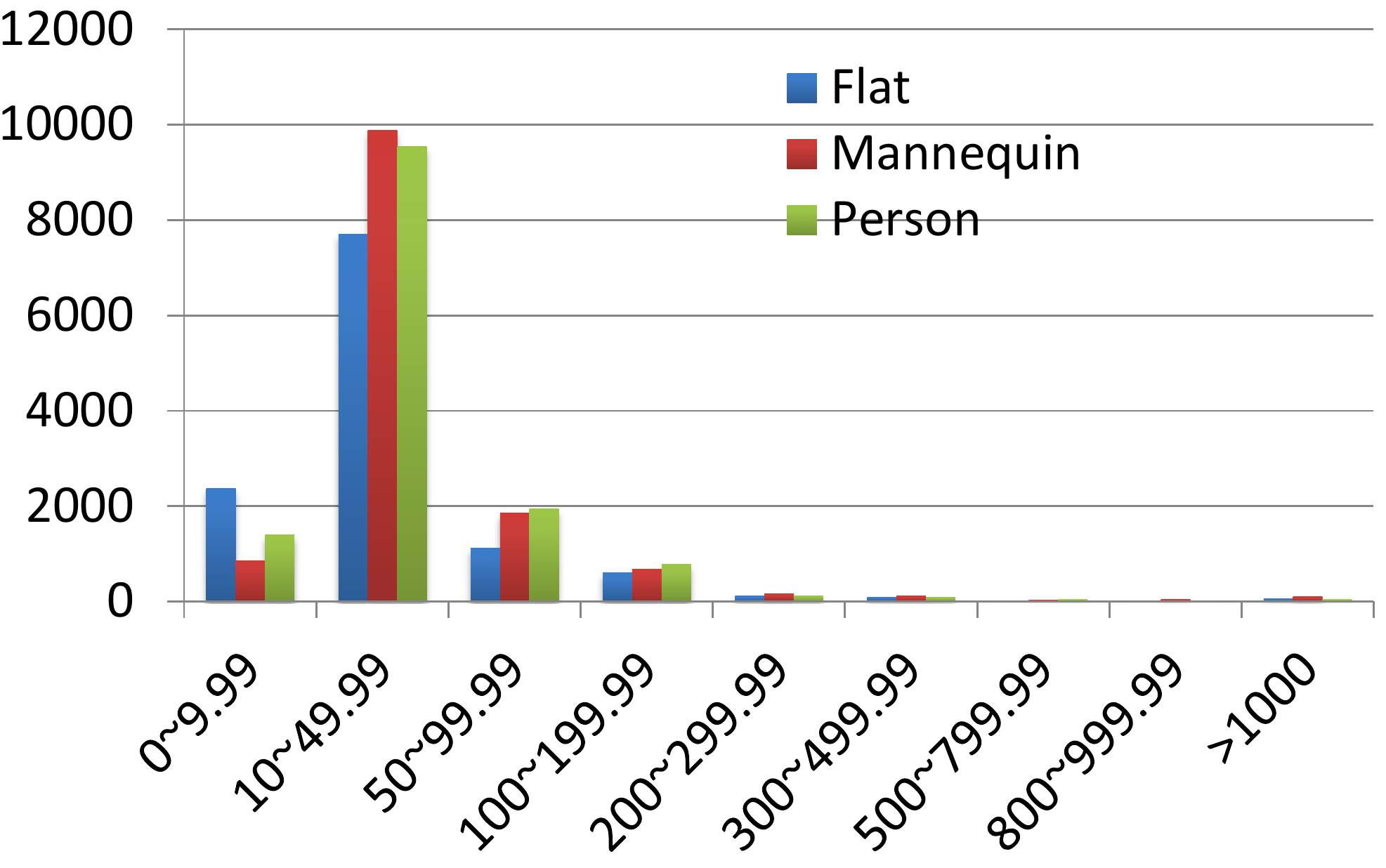,width=0.8\linewidth}  \\
{\scriptsize (b) Distribution of clicked items over various price segments} \\
\end{tabular} \vskip -9pt
\caption{Distribution of fixed-price items w.r.t. price segments. Dissected by price, user's preference on P-type is shown more clearer.} 
\label{fig:distri-price-seg}
\end{figure}

Is it possible that this shifting comes from other factors as aforementioned? To find out, 
we further investigate two additional important elements, i.e. price and seller type.
Price is often the driven element that influences decision making, especially given similar products.  
Figure~\ref{fig:distri-price-seg} shows the distribution of clicked and unclicked items, whereas buyers show strong inclination toward items presented in P-type even for different price segments. The results are generated using only fixed-price items, by which we exclude click actions that may be due to bidding on low price items. 

\begin{figure}
\centering \vskip -9pt
\begin{tabular}{c c c}
\epsfig{file=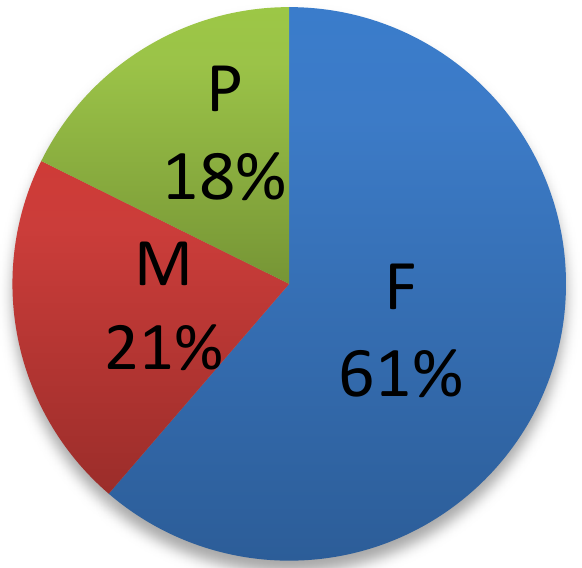,width=0.25\linewidth}  & &  \epsfig{file=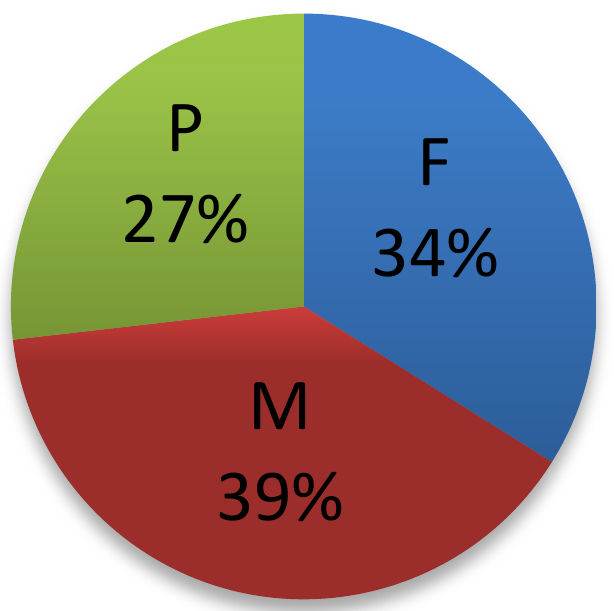,width=0.25\linewidth} \\
{\centering \scriptsize (a) Casual Seller Distribution} &  & {\scriptsize (b) Top Seller Distribution} \\
\end{tabular}
\begin{tabular}{c c c c c c c}
\multicolumn{3}{c}{\epsfig{file=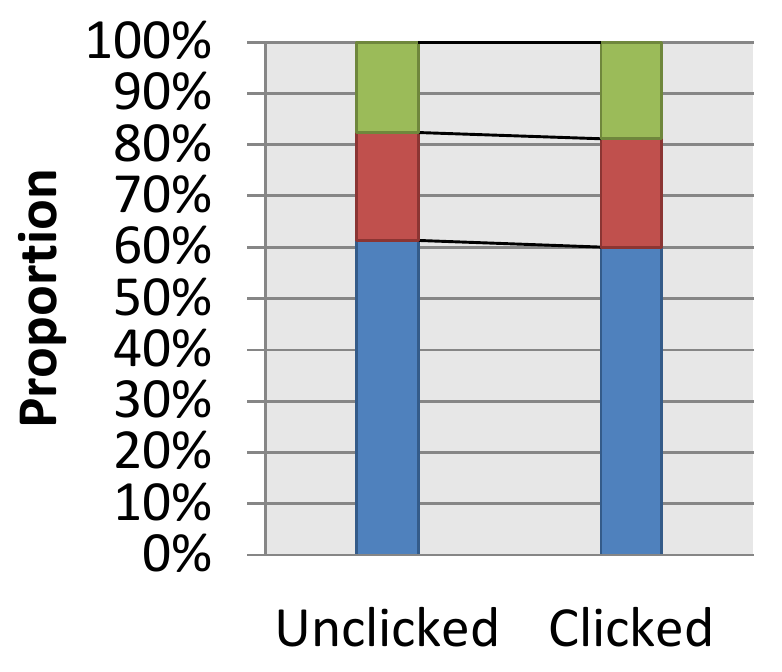,width=0.41\linewidth}}  & \multicolumn{4}{c}{\epsfig{file=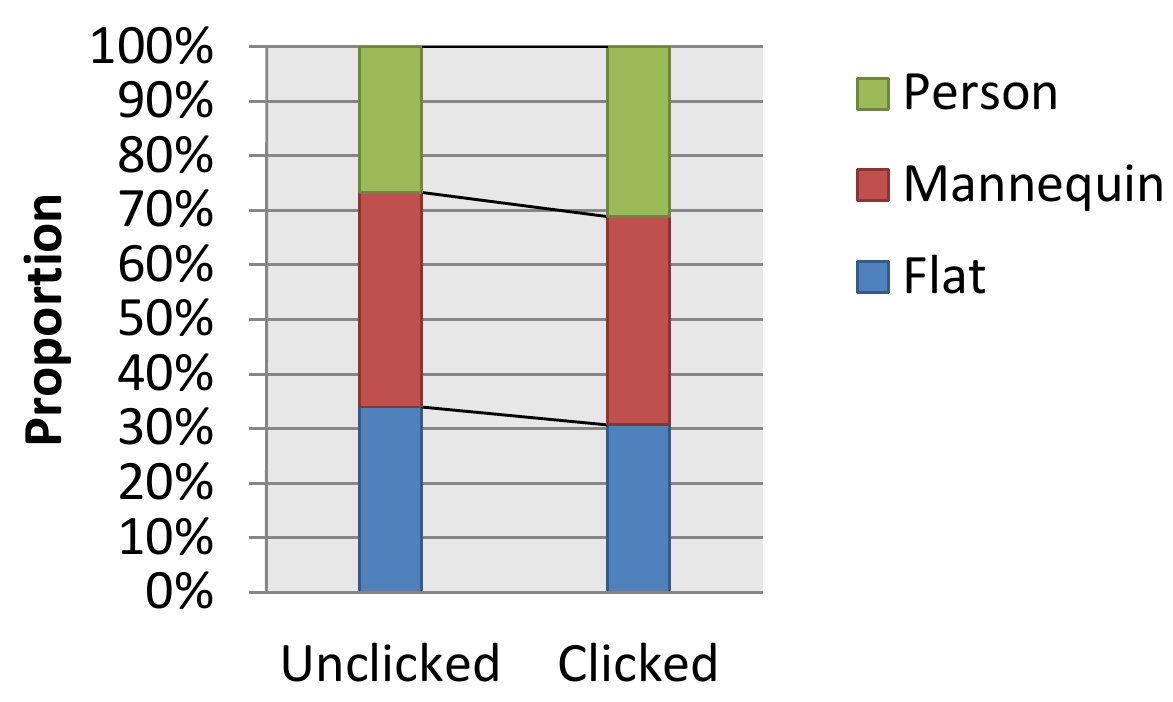,width=0.59\linewidth}} \\
\multicolumn{3}{c}{{\scriptsize (c) Casual Seller}}  & \multicolumn{4}{c}{{\scriptsize (d) Top Seller}} \\
\end{tabular}
\caption{{\fontsize{6.5}{8}\selectfont \textsf{PMF}} proportion for top sellers and casual sellers respectively(a,b), and the shifting before and after click(c,d). For both sellers, distribution shift indicates users favor P-type.}  
\label{fig:sellers}
\end{figure}

Seller type can be viewed as a historical indicator for trust worthiness of seller. Compared to casual sellers, power sellers often gain better trust because of the reputation accumulated through large amount of past transactions and customers.
It can be seen from Figure~\ref{fig:sellers} (a,b) that power sellers tend to use more P-type ($27\%$) to display clothing, which is $9\%$ higher than casual seller. This may be due to years of experience and better resources. Recall the distribution shifting in Table~\ref{tab:dist_shift_showed_clicked_unclicked}, it's worth to ask the question that is the result in Table~\ref{tab:dist_shift_showed_clicked_unclicked} due to that users chose more items from top seller? 
The answer is ``no''. As shown in Figure~\ref{fig:sellers}(c,d), for both sellers, distribution shift favors P-type. Changes in top seller is actually more significant with $4\%$ difference between unclicked and clicked items (from $27\%$ to $31\%$). 

\subsection{The Second Thought: Intention}
\label{subsec:watch-intention}

\begin{figure}[h]
\centering
\epsfig{file=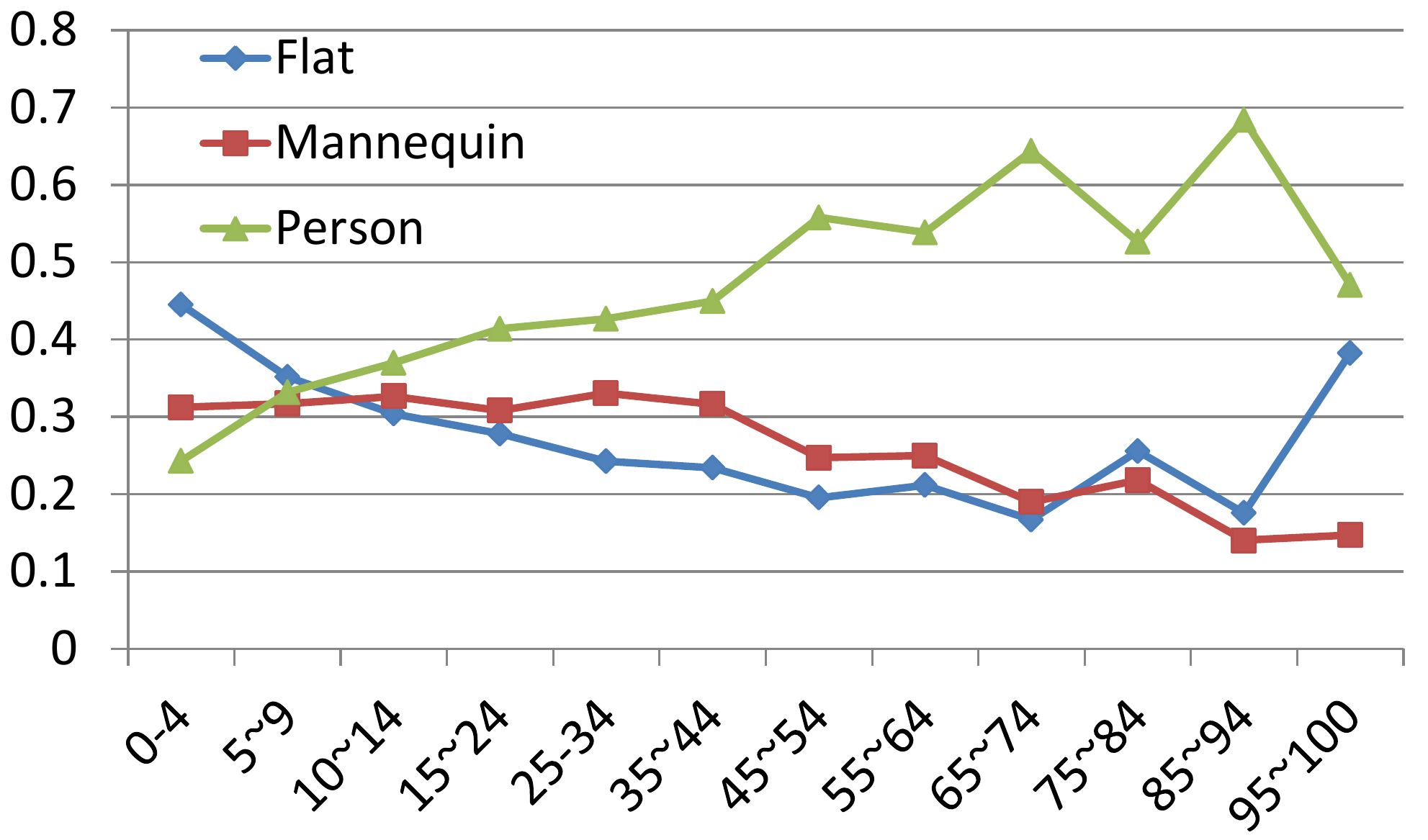, width=0.8\linewidth} \vskip -9pt
\caption{Normalized {\fontsize{6.5}{8}\selectfont \textsf{PMF}} distribution (y-axis) regarding to amount of ``Watch'' (x-axis). Proportion of P-type presentation increases for highly watched items as compared to less watched items.}
\vskip -10pt
\label{fig:watch_pmf_distri}
\end{figure}

Given higher attention drawn by P-type on the search result page, we are also interested in user actions on the view item page. Here we investigate buyer's ``Watch'' action, where user bookmarks the item for a more \textit{serious} evaluation, indicating more direct shopping intention. To dissect the influence, we compute the average watch count for each {\fontsize{8.0}{9.8}\selectfont \textsf{PMF}} type and for each seller group. Table~\ref{tab:avg-watch} and Figure~\ref{fig:watch_pmf_distri} suggest positive correlation of ``watch action'' with top seller as well as P-type product presentation. For items sold by either casual or top seller, P-type presentation helps increase the chance of being watched. Proportion of P-type image goes up for highly watched items as compared to less watched items.

\begin{table}[h] \small
\centering
\begin{tabular}{c | c | c}
	&  \multicolumn{2}{c}{Avg-Watch}  		\\ \hline
Type    		&\textbf{Casual-Seller}	&\textbf{Top-seller}	  	\\ \hline 
Flat			&1.48   			&1.89			\\
Mannequin		&1.89		&2.32			\\
Person			&2.73		&3.32			\\
\end{tabular}
\caption{Average ``Watch Count'' for each display type w.r.t. seller types. Results suggest P-type is correlated with higher average watch rate for both casual and top seller.}
\label{tab:avg-watch}
\end{table}

\subsection{The Ultimate Battle: Purchase}
\label{subsec:purchase}

Sell-through rate is the ultimate evaluation metric for an online listing. Table~\ref{tab:conversion-rate} lists the conversion rate of each display type ({\fontsize{8.0}{9.8}\selectfont \textsf{PMF}}), grouped by click action observed in the collected session data. Compared to unclicked items, clicked items show higher conversion rate, which is expected as user shows interest in the item through clicking, which leads to higher chance of purchase. Yet, most importantly, comparing the three types({\fontsize{8.0}{9.8}\selectfont \textsf{PMF}}), items displayed in P-type shows better sell-through for either clicked or unclicked items.

\begin{table}[h] \small
\centering
\begin{tabular}{c | c | c}
	 		&\textbf{Clicked Items}	&\textbf{Unclicked Items}   \\ \hline 
Flat		&41.88\%		&26.05\%  \\
Mannequin	&42.45\%			&23.46\%  \\
Person		&47.94\%			&28.23\%  \\
\end{tabular}
\caption{Conversion Rate for three display types for clicked and unclicked items in the collected session data, where items displayed by P-type show better sell-through rate.} 
\label{tab:conversion-rate}
\end{table}

\section{True Voice Inside User}
\label{sec:user-study}

As demonstrated in previous sessions, evidences from the real-world online marketplace data show that shoppers have different preference on different display types. Yet, deeper questions still exist. 
For example, by quantizing attractiveness using three common clothing display types, given the huge favor of P-type, a natural question is whether consumers really think such type is most attractive. In other words, we would like to verify our hypothesis that there is distinct attractiveness difference between the three PMF types that causes user taking different choices.
Another question comes from the concern of non-visual factors that may have affected user's choice. 
Although random sampling of data may help smooth out some of the hidden factors, given the complexity of the real marketplace, we still need to ask: could the observed bias be a causal effect or correlation effect by unknown variables, rather than the perceived visual effects? For example, textual meta information associated with the fashion product, i.e. price, brand, seller reputation, shipping cost and time, could have all impacted user in their logic process of browsing and purchase.
Hence, we are interested in investigating user response by isolating visual information. We would like to know is there any consistence or discrepancy in consumer's preference between a simulated environment and a real marketplace.

\subsection{Experimental Design}

To do so, we randomly select $4968$ images with $1656$ per PMF class from the previous collected data, and conduct a user study through Amazon Mechnical Turk. 
In order to avoid putting any bias to the user, we did not reveal the true purpose of the study. There is no direct question asking user's preference on the product presentation, or clothing display type, such as human model, mannequin or flat. Instead, we only ask user to freely rate the images based on their personal preference. 
We show in total 9 images together for each MTurk worker (per HIT). They are randomly selected from each class (3 per class), and shuffled to mix their order of presenting. 
None of the meta information is given.
The worker's task is to choose the best (like) or the worst (dislike) items based on what they see in each small image group, and make choices for three different scenarios: buying, sharing and gifting, as shown in Table~\ref{tab:three-senarios}.

\begin{table}[h] \small
\centering
\begin{tabular}{p{0.03cm}p{3cm}|p{4.6cm}}
 &\textbf{Scenario}	&\textbf{Task}  \\ \hline 
I.	&Buying For Yourself				& select two best (like, most likely to buy) items and two worst (dislike, least likely to buy) items based on your preference \\
II.	&If You Share					& select multiple items you would like to share through any social networks (\textit{NOT buying}), e.g. facebook, twitter \\
III.	&Buying as a Gift to Others		& select multiple items you would like to buy as a gift to others  \\ 
\end{tabular}
\caption{Worker are given task on three scenarios to freely pick the images based only on visual information.} 
\label{tab:three-senarios}
\end{table}

From user's feedback on the first task (buying scenario), we categorize product images into tree groups: \textit{Like}, \textit{Neutral} and \textit{Dislike}. The two best images identified by the user are categorized into ``\textit{like}'', and the two worst images are labeled as ``\textit{Dislike}''. To understand the reason behind user's action of selecting the best two images (``like'' group), we further ask the same user to choose multiple reason from nine given options if any applies and vote for the top ones. They are also allowed to freely type in their own reasons. The nine candidate reasons are:

\begin{itemize}
\item Visually attractive
\item Clothing looks high quality
\item Clothing is unique
\item Style fits for me
\item Fit for my purpose (e.g party or casual wearing)
\item Shows details of the clothing
\item Image quality is better
\item Feels less risky
\item Good add-on for what I've own already
\end{itemize}

As mentioned before, image is only one type of information that during the shopping process consumers are collecting and evaluating. The final decision is made by carefully examining all available factors. Preference or bias from user may also exist in other dimensions, such as brand and price, which are often recognized as two major influential factors. Therefore, we are interested in how consumer reacts towards these two primary aspects. During the experiment, we asked the user with two additional questions: 

\begin{itemize}

\item \textbf{Price}: We asked the user whether he/she would like to change their choice given the condition that the \textit{Like} group that they selected has higher price, whereas the \textit{Neutral} group has lower price. Specifically, given a range of price differences, how much price advantage the \textit{Neutral} group needs to have so as to make the user to switch from their initial choice which was purely made based on only visual information?

\item \textbf{Brand}: We asked the user whether he/she would like to change their choice given the condition that the \textit{Like} group they chosen are non-branded, whereas items from the \textit{Neutral} group are of famous brand. Given the same price, would the user switch his/her choice from \textit{Like} to \textit{Neutral}?  

\end{itemize}

Overall, we collected more than 500 HITs from various users. There are 254 hits from female users, and 285 from male users. About 52.38\% are from the age group of 20-30, and 39.38\% from the age group of 30-40. We also manually removed some non-eligible data if we notice that Turkers submitted incomplete or incoherent work.

\subsection{Implications \& Discussion}

\begin{figure}[h]
\centering 
\epsfig{file=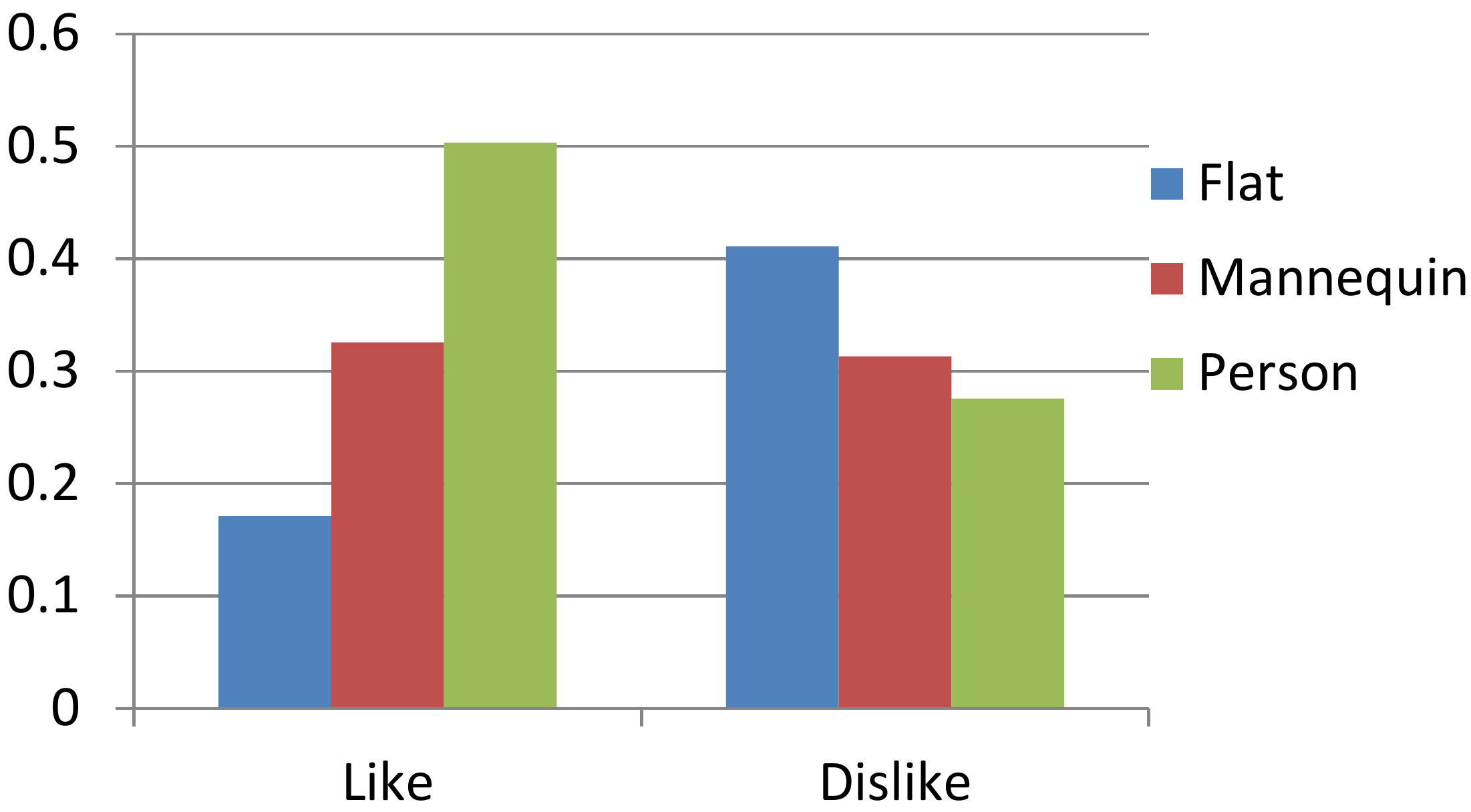,width=0.68\linewidth}  \vskip -9pt
\caption{Results from task I (Buying for yourself). Consumers are asked to select two \textit{Like} and two \textit{Dislike} items based on only visual information from the 9 items shown together. The height of bar indicates the proportion of each PMF type in each group.}  
\label{fig:good-bad-distribution}
\end{figure}

From the collected data, it would be interesting to see that what are the items users \textit{like} and what they \textit{dislike}, and whether there is any connection with the PMF model. 
Figure~\ref{fig:good-bad-distribution} shows the results for Task I (buying for oneself). Consumers are asked to select two \textit{Like} and two \textit{Dislike} items based on only visual information from the 9 images shown together.
Results are clear and as expected. For images that users label as \textit{Like} - most likely to buy, we can see strong favor towards P-type images (P>M>F). 
For images people think as \textit{Dislike} - least likely to buy, the proportions are apparently in a reverse order (P<M<F).
Overall, results are consistent with the observation from the real online market but with noticeable distinction between M and F-type. Given the noises in online data, we believe such results is a better reflection of the genuine preference of users if only visual information is given.


\subsubsection{Reasons: The logic behind the scene}

\begin{figure}
  \centering   
  \begin{overpic}[scale=0.32]{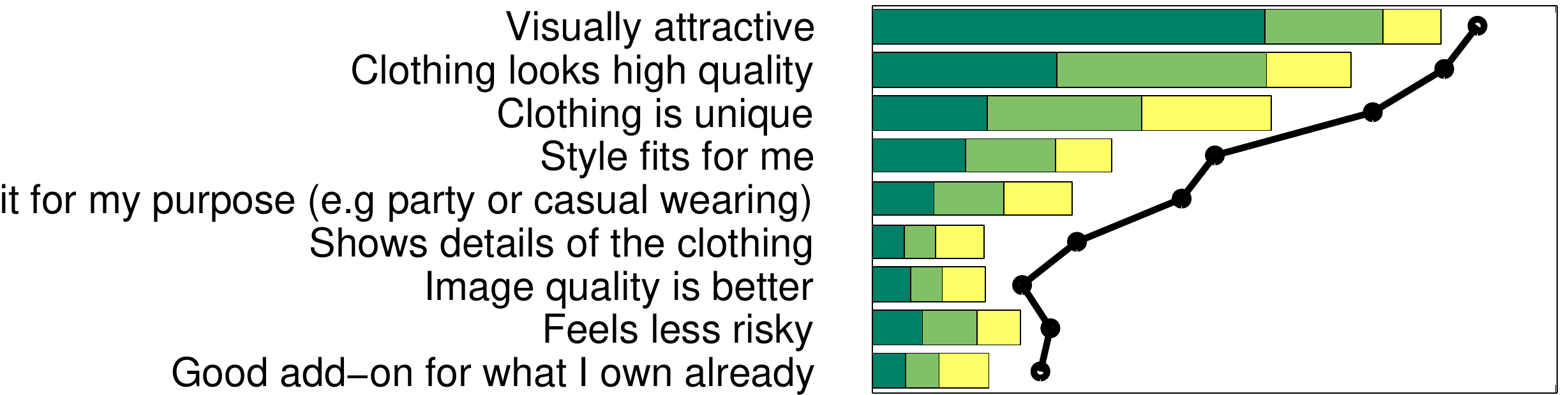}
     \put(80,3){\includegraphics[scale=0.21, angle=90]{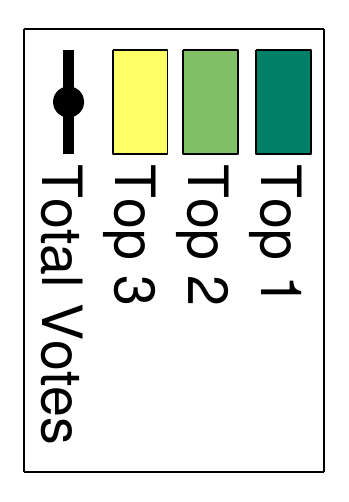}}  
  \end{overpic}  
\caption{Total votes for each reason, and choices for top reasons.} \vskip -9pt
\label{fig:reasons-all-each}
\end{figure}

Our goal is to further analyze user's behavior to ascertain its possible causes, i.e., what drives user to make their choice? Is there distinct attractiveness difference between the three PMF types that causes user favoring more on P-type presentation, and favoring least on F-type presentation? In the survey, we provide nine candidate options for each worker on why they \textit{like} certain items and most likely to buy. We are interested in not only knowing what matters, but also what matters most, i.e. what are the top reasons that affect people's decision.
Figure~\ref{fig:reasons-all-each} shows the overall votes for each reason (black curve with dot marker), and the votes for being the top-3 (colored and shown in bar plot). They are ranked based on the overall number of votes.
Among the 9 given reasons, ``\textit{Visually attractive}'' appears to be the most important reason, followed by ``\textit{Clothing looks high quality}''. However, although these two have close amount in terms of overall votes, ``\textit{Visually attractive}'' is voted twice as much as  ``\textit{Clothing looks high quality}'' as the top-1 choice.
On the other end of spectrum, compared to the last 3 reasons,``\textit{Shows details of the clothing}'' has higher number in terms of overall votes, but it is the least voted one to be in the top-3 among all nine options. User seems to concern more about the perceived risk (reason ``\textit{Feels less risky}'').  

\begin{figure}[h]
\centering 
\epsfig{file=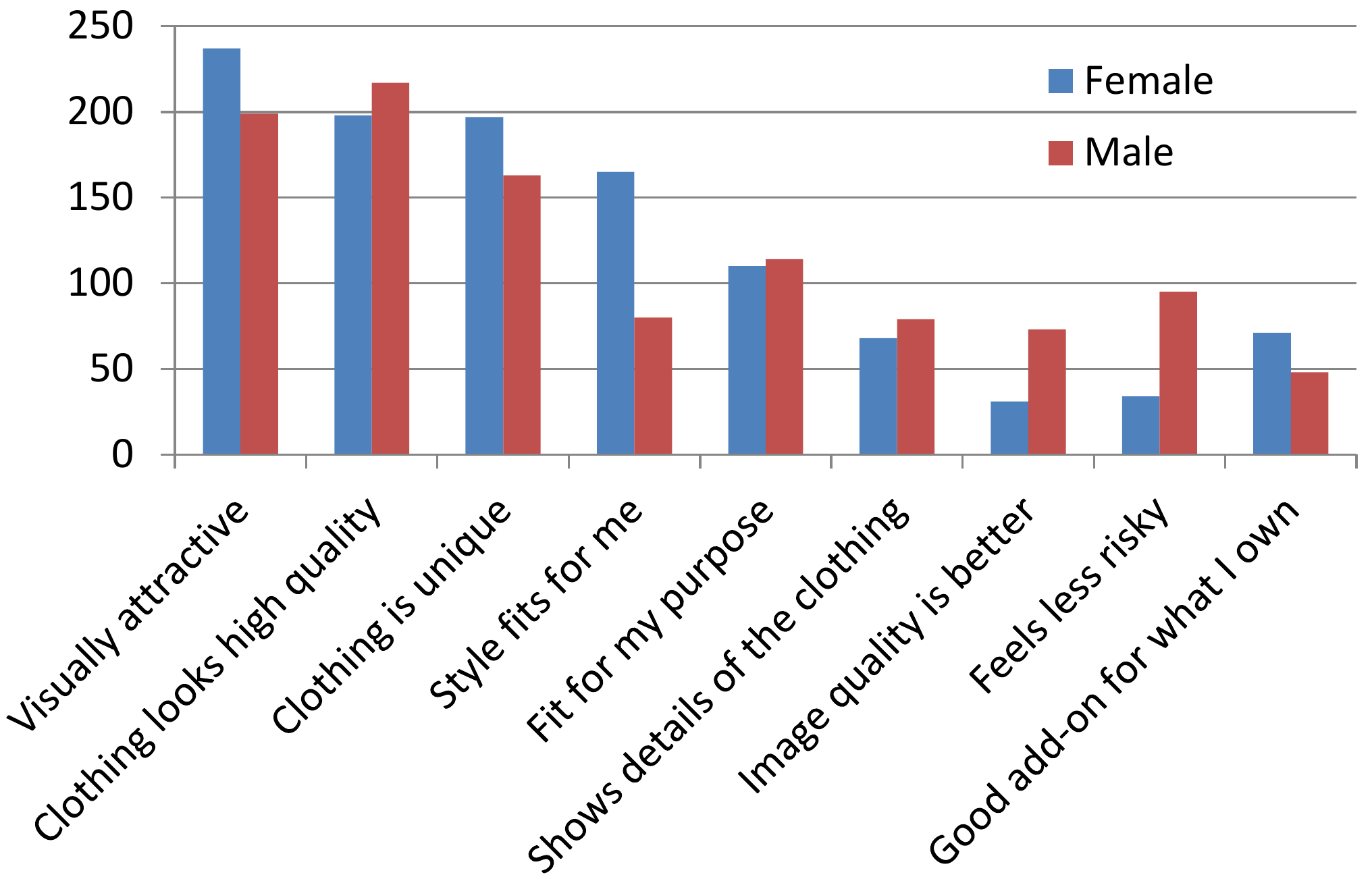,width=0.8\linewidth} 
\caption{Difference in choices of reasons w.r.t gender.}    \vskip -9pt
\label{fig:reason-differ-in-gender}
\end{figure}

Is there any difference between the reasons chosen by female or male subject? As shown in Figure~\ref{fig:reason-differ-in-gender}, for most of the cases female and male subjects are consistent with each other. However, they deviate on three major points: ``\textit{Style fits for me}'', ``\textit{Image quality is better}'', ``\textit{Feels less risky}''. 
The first difference in ``\textit{Style fits for me}'' is reasonable as female user tends to be more critical on style as they are probably imaging they themselves are wearing the dress. In contrast, male subjects possibly only give opinions according to their aesthetic standard. Higher votes from male subjects on reasons ``\textit{Image quality is better}'' and ``\textit{Feels less risky}'' suggest that man seems to be more objective in this case by emphasizing more on cost and risk, whereas women on the other hand are less worried about external factor other than the clothing itself.

The survey also gets many direct inputs from the user. Some of the reasons addressed by the users are very specific. Although we did not disclose the true purpose of this study, there is one particular user specifically mentioned that ``\textit{The dresses that were on models were easier to assess}''. For other reasons, we group them into few categories as shown in Table~\ref{tab:input-reason}. Most of these reasons are related to specific \textit{attributes} of the clothing, such as style, length or material. That said, although this study focuses on the presentation of the product, there is no doubt that quality of the product itself is one of the primary factors affect people's decision. We expect such insights from users can shed lights for future study on semantic attribute based clothing categorization for promoting personalization in fashion e-commerce.

\begin{table}[h] \small
\centering
\begin{tabular}{p{1cm}p{5.5cm}}
\textbf{Group}	&\textbf{Reason Statements}  \\ \hline 
color     & The color is lively and vibrant.\\
		  & It looks like it will never fade. \\  
		  & The color black is a safe and classy so it still trendy over time.  \\ \hline
material  & I like nice and smooth fabric \\ \hline
length    & The hem is not too short and not too long.  \\  \hline
fashion   & The dress suits for petite women. It will make you sexy.  \\
		  & Regardless of what you wear just top with that blazer and you look fab. \\ \hline
style     & The length and color affected my choice, but mostly the style.  \\
		  & I like it looks simple but elegant.  \\
		  & The dress has sophisticated looks but still classy.  \\
		  & The styles of the dress is not too revealing but still sexy if you already wear it. \\
		  & I mainly chose the dresses I did as a result of what I imagined my girlfriend/significant other wearing, what I'd find visually appealing, and the styles. \\
		  & classic styles, I like the one with jacket \\ \hline
fit       & my choice depends on which i like and which will fit for me. \\ \hline

\end{tabular}
\caption{Reasons collected directly from user input.}  \vskip -9pt
\label{tab:input-reason}
\end{table}

\subsubsection{What can make you switch choice?} 
The previous results provide some interesting insights that tie back to our initial goal of segmenting influences of visual presentation from other meta information. 
Consider a real shopping process where many related information of the product are evaluated, we are particular interested in two primary aspects: price and brand. Without being very specific, we question in the survey that whether user would like to change their initial choice (based on only visual information) when knowing additional price and brand information. We only consider the item group marked as \textit{Like} and \textit{Neutral}. For price, we assume that there is an additional cost of the \textit{Like} group as compared to \textit{Neutral} group. For brand, we test the user by indicating items they \textit{Like} are un-branded, while items in the \textit{Neutral} group are from top brands.
Table~\ref{tab:price-brand-impact} shows that users are more likely to switch if there are price differences. But surprisingly, majority users tend to stick to their choice regardless of the brand, which is contradict to our expectation. This is because brand is often thought to play an important role in affecting people's decision in fashion domain.

\begin{table}[h] \small
\centering
\begin{tabular}{c | c } 
\textbf{Price Differ}:	&\textbf{Votes for Switch}   \\ \hline 
0-10		&35		  \\
11-20		&113		 \\
21-30		&123		 \\
31-40		&83		 \\
41-50		&64		  \\
>50			&128		  \\	\hline
\textbf{Brand Switch}: &\textbf{Votes}		  \\	\hline
Yes		&193		\\
No	&343		\\
\end{tabular}
\caption{The price \& brand impact. For price: table shows additional cost of items in \textit{Like} group as compared to items from \textit{Neutral} group and the votes for a switch at each price range. For brand: assume items in \textit{Like} group are non-branded, whereas items of \textit{Neutral} are from top brand.} 
\label{tab:price-brand-impact}
\end{table}

\subsubsection{Sharing \& Gifting: because you like?}
\begin{figure}[h]
\centering 
\begin{tabular}{c}
\epsfig{file=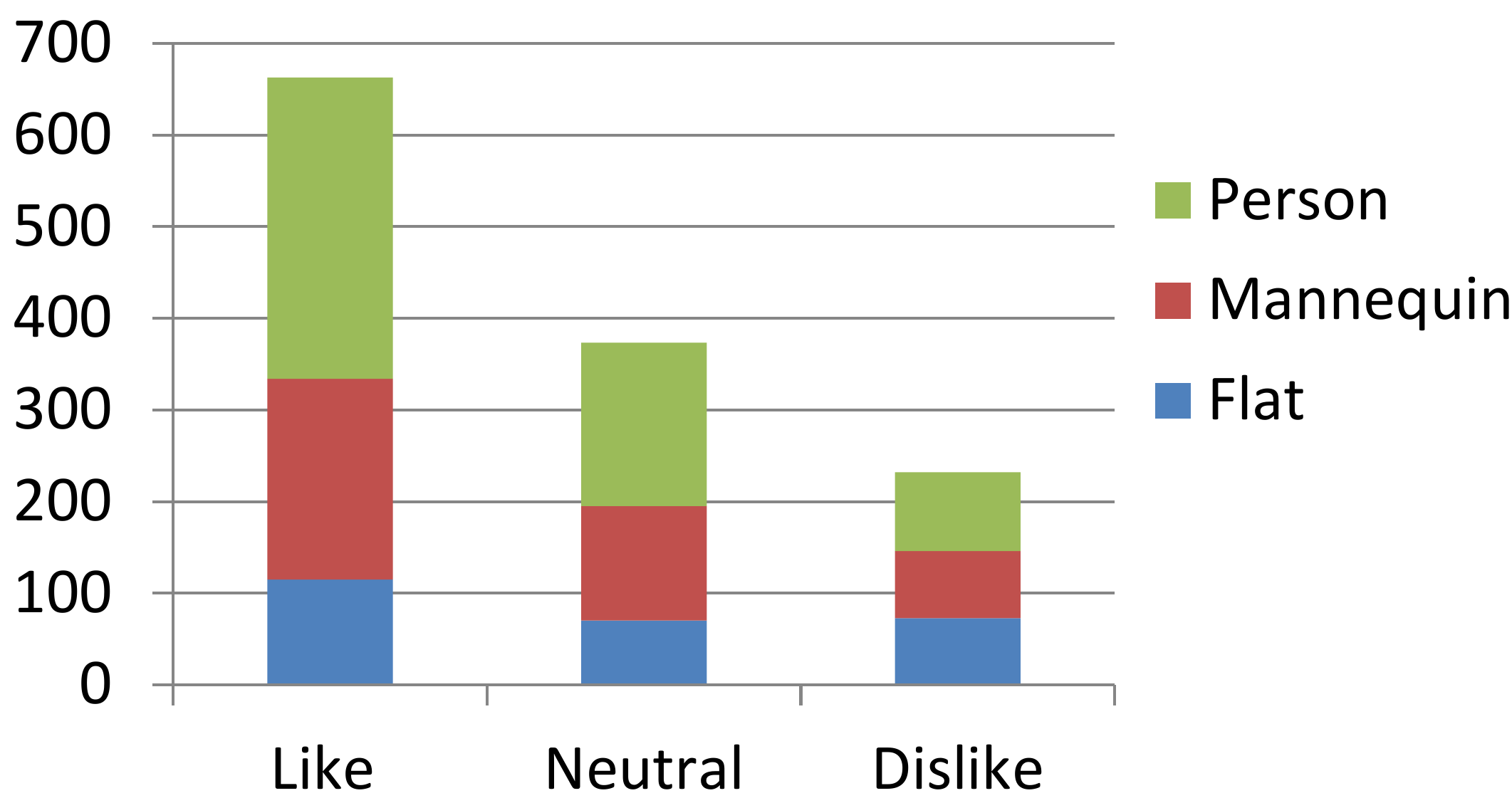,width=0.7\linewidth}  \\
{\small (a) Distribution of shared items} \\
\epsfig{file=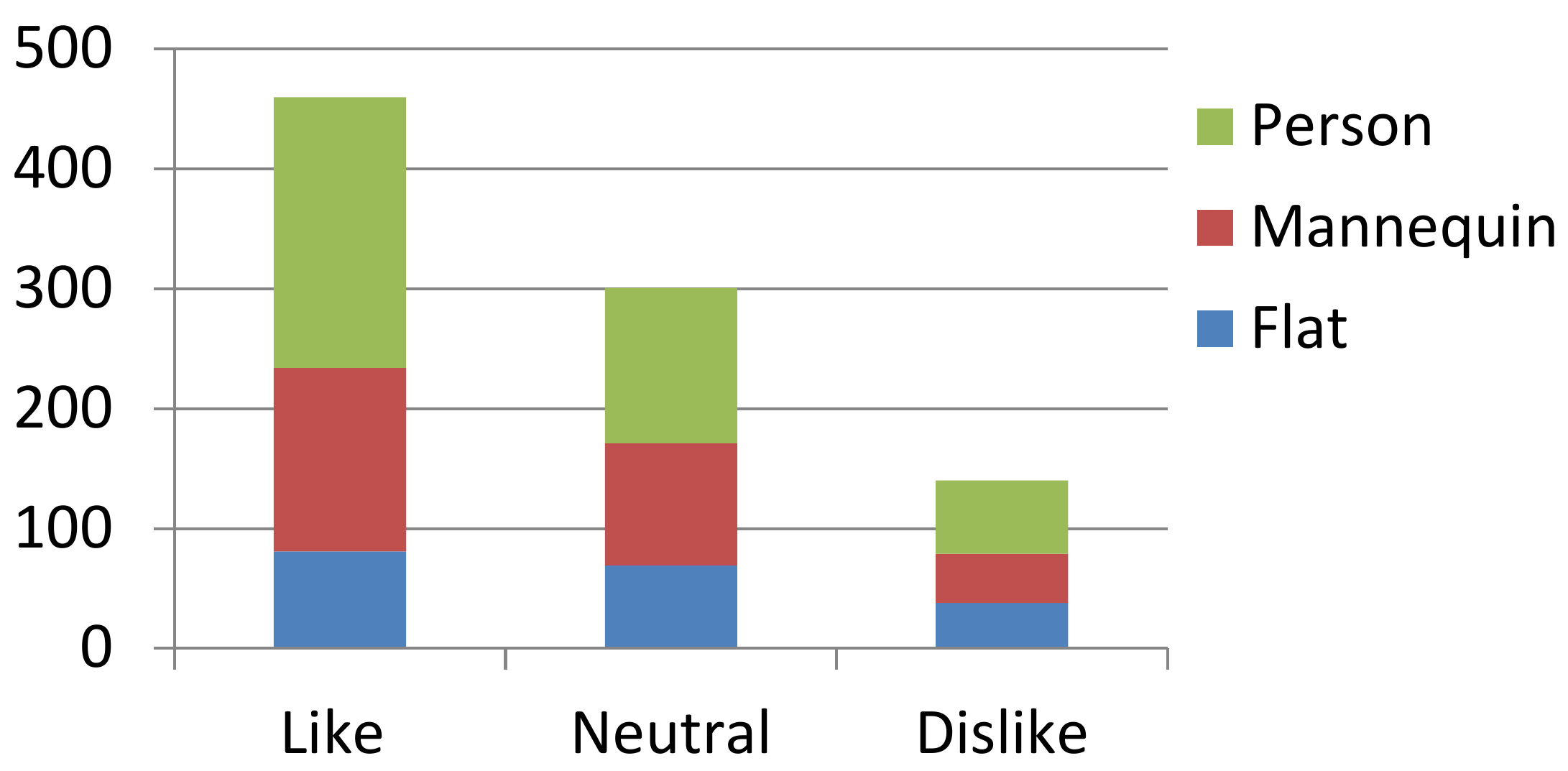,width=0.75\linewidth}   \\
{\small (b) Distribution of gifted items}  \\
\end{tabular}
\caption{Distribution of shared \& gifted items. While people tends to share or gifting more for what they like, given the changes in the context and targeted subjects, there is a noticeable discrepancy between what are liked and what are shared or gifted.}   \vskip -9pt
\label{fig:sharing-gifting}
\end{figure}

In terms of sharing and gifting, we found that users are consistently sharing or gifting more for the items that they \textit{like}, i.e. more likely to buy for themselves, as shown in Figure~\ref{fig:sharing-gifting}. As expected, items displayed using human model are more likely to be shared and gifted. 
However, given the fact that there is small portion of items for which people dislike but are willing to share and gifting, it means that the motives of sharing or gifting is more complex that cannot be simply explained by personal favoring or not. As the targeted subject has changed from oneself to a lager audience in the social network or people to gift with, preferences are catered to different domains. This can be explained by one user's feedback on her choice: ``\textit{For myself I would pick classic styles, but if I was giving to my young niece I would choose something more fun or cute}''. 

Another natural question is whether sharing and gifting share some common properties. 
To analyze how much these two sets overlap (sharing \& gifting), we compute the overlapping score by:
\begin{equation}
\zeta(\Omega_s, \Omega_g) = \frac{  \vert \Omega_s \cap \Omega_g \vert }{\min (\vert \Omega_s \vert, \vert \Omega_g \vert  )  }
\end{equation}
\noindent where $\Omega_s$ and $\Omega_g$ are sharing and gifting item set, respectively.

\begin{figure}[h]
\centering 
\epsfig{file=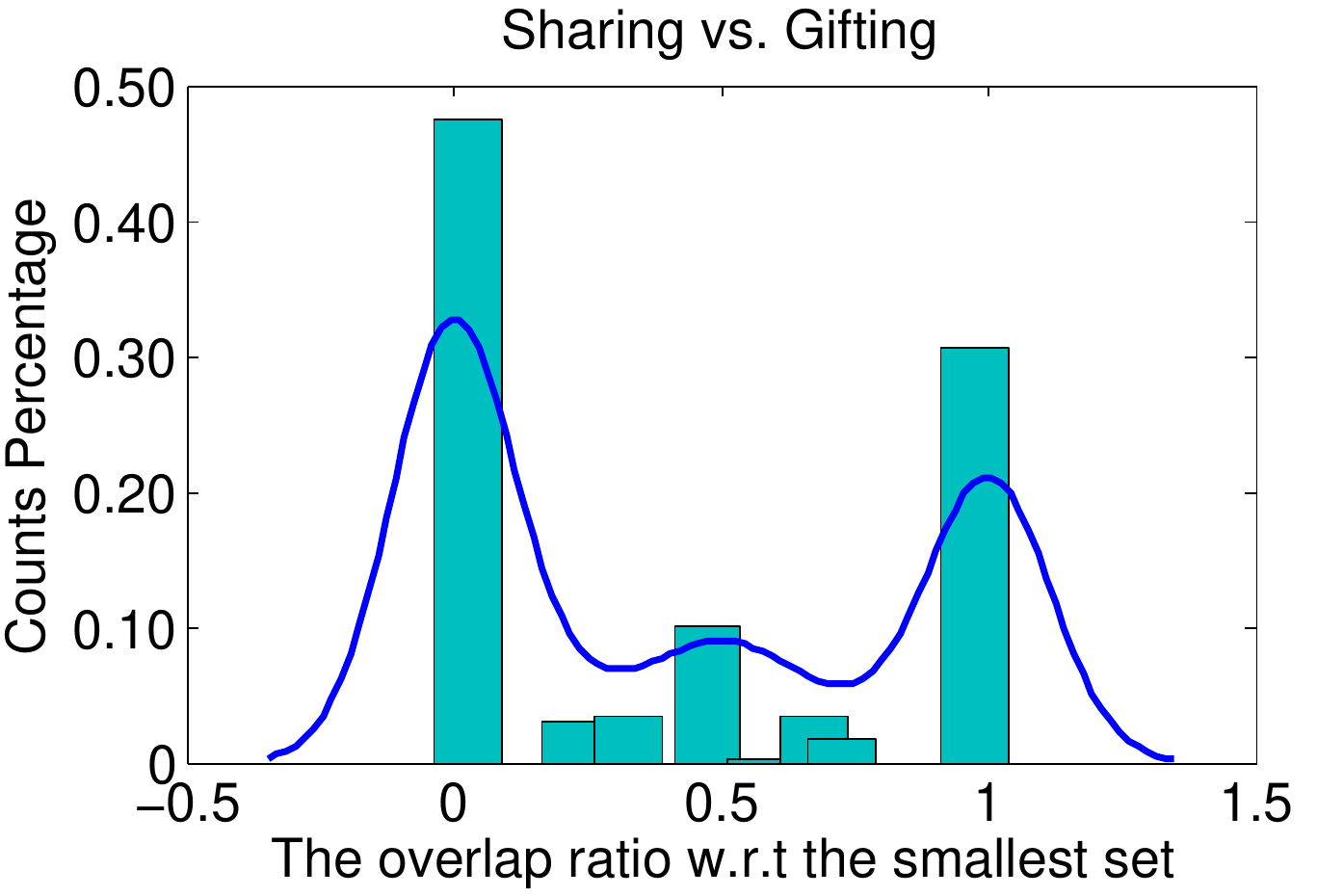,width=0.6\linewidth} 
\caption{Evident differences can be seen in choices of sharing and gifting, where close to 50\%, there is no overlap between the sharing and gifting. This is possibly due to the differences in behavior motives and targeted group.}  \vskip -9pt
\label{fig:differ-in-sharing-gifting.}
\end{figure}

Clearly, results in Figure~\ref{fig:differ-in-sharing-gifting.} indicates that there are significant deviation between sharing and gifting. Close to 50\%, there is no overlap between the sharing and gifting. This provides an interesting cue that behavior changes in different contextual are not neglectable and deeper understandings are needed before leveraging social context for e-commerce recommendation or promoting consumer engagement.

\section{Weighing User Preference}
\label{sec:weighing}

By categorizing attractiveness into a three-point scale using existing clothing presentation, evidences from previous section have shown that users are more drawn to P-type clothing presentation. We would like to ask: is there significant differences between the preference of each attractiveness level, i.e. each type, and how much? Is there a quantitative way to compare them? Here we propose a {\fontsize{8.0}{9.8}\selectfont \textsf{PMF}}-user choice model to quantify user preference. We hope this quantitative study can help e-retailers to have a clearer idea of the level of risk and choose the right presentation type. 

In both the experiments conducted on the real online marketplace or on the designed user subject study, at each time, users are shown multiple items/images simultaneous from the inventory. The proportion of each type of image might be biased, for example in the real online marketplace, or non-biased, for example in our user study experiment. The click or selection choice made by user is affected by both the amount/proportion that each type is showed, and by consumer's own preference. Our goal is to estimate the level of preference by utilizing the click data which is affected by such bias toward visual presentation. 

In fact, a similar sampling problem has been studied in statistics, namely urn model~\cite{fog2013biased}, whereas multiple color balls with different weights and amount are considered. Often a hypergeometric distribution, which is the discrete probability distribution, is used to model the distribution generated by picking colored balls at random from an urn without replacement. It has been widely used in many areas, such as information retrieval, social mining~\cite{tsagkias2011hypergeometric}. Various generalizations to this distribution exist for cases where the picking of colored balls is biased so that balls of one color are more likely to be picked than balls of another color. This is often refereed to as \textit{Noncentral hypergeometric distributions}. 

Therefore, we employ \textit{Multivariate Fisher's noncentral hypergeometric distribution} (MFNCHypergeo) to model user preference on each attractiveness level represented by different display types. 
MFNCHypergeo is a generalization of the hypergeometric distribution where sampling probabilities are modified by weight factors. It can also be defined as the conditional distribution of two or more binomially distributed variables dependent upon their fixed sum. 
To use this model, we make the following assumptions:
\begin{itemize}
\item We assume that each item is taken from a finite source containing different kinds of items without replacement.
\item Items are taken independently of each other. Whether one item is taken is independent of whether another item is taken. 
\item The probability of taking a particular item is proportional to its ``\textbf{weight}'', which in our case refers to the PMF preference level by user. The weight or preference of an item depends only on its kind.
\end{itemize}



Assume there are total $\mathbb{C}= \lbrace 1,...,c \rbrace $ kinds of items, where in our case $\mathbb{C}=\lbrace p,m,f \rbrace$. 
Let $\gamma_i$ denote the initial number of items of each kind and $\eta_i$ be the number of items of each kind sampled (measured by click or select). $n=\sum_{i=1}^{c}\gamma_i$ is the total number of items before sampling. $\widehat{n}=\sum_{i=1}^{c}\eta_i$ is the total number of items sampled without replacement 
in such a way that the probability that a particular item is sampled at a given draw is proportional to a property $\mathcal{W}_i$, which is the weight or odds of  an item depends only on its kind. 
The items have different weights which make the sampling biased in favor of the ``heavier'' items. Here, the weight represents the level of \textit{preference} toward each type, or attractiveness level. The smaller the original proportion $\gamma_i$ before sampling is, the harder items of type $i$ show up in search result and picked by user. The higher preference $\mathcal{W}_i$ is, more likely type $i$ will be selected. Thus, $\eta_i$, which is the post distribution of type $i$ is affected by both factors: distribution bias represented by $\gamma_i$ and preferences bias $\mathcal{W}_i$.
The probability mass function of Fisher's noncentral hypergeometric distribution is given by
\begin{align}
\text{dMFNCHypergeo}(\vec{\eta}; \vec{\gamma}, n, \vec{\mathcal{W}}) &=  \frac{ g(\vec{\eta}; \vec{\gamma}, n, \vec{\mathcal{W}})}{ \sum_t \in \Xi  g(\vec{\eta}; \vec{\gamma}, n, \vec{\mathcal{W}} )}  \\
\text{where} ~ g(\vec{\eta}; \vec{\gamma}, n, \vec{\mathcal{W}}) &= \prod_{i=1}^{c} \left( \begin{array}{c}
																		\eta_i \\
																		\gamma_i \end{array} \right)\mathcal{W}_i^{\gamma_i}  \\
\text{and domain} ~\Xi = \lbrace  \vec{\eta} \in \mathbb{Z}^c \mid \sum_{i=1}^c \beta_i &= \widehat{n} \wedge  \forall i \in [1,c]: 0 \leq \eta_i \leq \gamma_i  \rbrace
\end{align}


\begin{table}[t] \small
\centering 
\begin{tabular}{c | c | c }
\textbf{Type}  	& \textbf{D1} & \textbf{D2}  	\\ \hline
Flat		 		&0.8566		& 0.2341 \\
Mannequin		&0.8609		& 0.5217 \\
Person			&1			& 1  	 \\
\end{tabular}
\caption{Estimated preference level $\mathcal{W}_i$ of each {\fontsize{6.5}{8}\selectfont \textsf{PMF}} type by the proposed {\fontsize{6.5}{8}\selectfont \textsf{PMF}} User Choice models on two dataset. D1: data from the real online market. D2: data from the user survey.} 
\label{tab:est-weights}
\end{table}

We apply this model to the data curated from both online market place and our user group study.
For online market data, we model the sampling process by search click data from each search session as anlayzied in section~\ref{subsec:fist-glance}. For user group study, we only utilize the data from task I (buying for oneself). The sampling process is defined by selecting the items that user \textit{likes}, i.e. the post-distribution $\eta_{i\in(p,m,f)}$ is the distribution of each PMF type among the items marked as \textit{Like}.

Table~\ref{tab:est-weights} lists the predicted preference level from these two types of data, subject to $\mathcal{W}_p=1$ given that weights in MFNCHypergeo can be arbitrarily scaled. For both data, P-type clearly gains the highest preference. Results from survey data show that difference between M-type and F-type are quite significant. However, such difference is hardly observable from the real online market data. 
There are few possible reasons. First, we found that in the online market data {\fontsize{8.0}{9.8}\selectfont \textsf{Flat}} category consists of many non-dress items, which are retrieved when user uses queries like ``\textit{black dress shoes}''. E-shoppers may tend to click those {\fontsize{8.0}{9.8}\selectfont \textsf{Flat}} items either because they are exploring, or searching for coordinate items (shoes, belts, etc) that match well with black dress. Second, it is also possibly because that other dimensions of the product affect user's choice as various meta information such as seller types, shipping, brand and price are shown to user. This further validate the necessity of conducting a user survey in order to ascertain our hypothesis.
Finally, although the use of {\fontsize{8.0}{9.8}\selectfont \textsf{Mannequin}} enables user to imaging the volume of the clothing and the feelings of clothing be wearing, it is an inanimate human-size figure. Thus it's hard to arouse pleasing feelings or interest, from viewers.

\section{Search Reranking}
\label{sec:search-reranking}

In typical search systems, given a query or a statement of information needed, the task is to estimate the relevance score $R(\mathbf{x})$ of each items in the inventory and return in the order by their relevance score, where $\mathbf{x} \subset \mathbb{X} $ represents the item.
In literature, many approaches have been proposed for reranking, ranging from sophisticated fusion of multiple modalities to incorporating user feedback information. 
Some approaches rely on exploring visual features for better matching to improve the search relevance, whereas 
image relevance serves as a conditional variable in $P(Y|X)$, where $Y$ is a random variable representing search relevance. The posterior probability is then used for $R(\mathbf{x})$ in ranking.

Given the observation from previous sections, we believe there are potential benefits to incorporating visual attractiveness into reranking schema, for promoting not only what users ``\textit{want}'', but also what they ``\textit{like}''.
This is particularly true for fashion e-commerce. Given the fact the fashion is a particularly visual oriented field, visual information, not only the image content, but also the way that they are presented are highly important to have effective communications.
Therefore, we propose a new visual attractivness based reranking schema incorporating both user preference and variables representing the presentation efficacy. The proposed ranking approach accounts for a subtle but important difference between conceivable alternatives with respect to image content and preference from user on perceived visual effect.

\subsection{Visual Attractiveness Re-ranking Model}

We formulate the ranking problem as follows: 
As user preference often captured by click-data, we assume that we have a set of $N$ ranked retrieval results \allowbreak $\Theta= \lbrace (\mathcal{X}, r^*, \mathcal{Y}): (\mathbf{x}_i, d_i, y_i)_{i=1, \ldots, \mathcal{T}}^j \rbrace$ where $j=1 \ldots N$, and for each group we only focus on top $\mathcal{T}$ items. 
We assume that the top $\mathcal{T}$ items are all highly relevant as our goal is to learn a re-ranking function the refine such ranking by taking account presentation efficacy for what people \textit{like} in addition to what they \textit{want}. 
$\mathbf{x}_i$ is the image descriptor (Bow) defined in section~\ref{sec:vision}, $y_i \subset \mathit{Y}= \lbrace 0,1 \rbrace$ is the ``relevance'' label, captured via click-data. $y_i=1$ indicates user selects the given item, while 0 otherwise. $d_i$ denotes a list of ``order preferences'', which is the integer indicating its order. For example, if $d_i > d_j$, the item $i$ should be ranked higher than item $j$. In our set up, we group this preferences into three levels: $ d \subset \lbrace 1, 0 -1 \rbrace$.  
For the top relevant items $\mathcal{X} =( \mathbf{x}_1, \mathbf{x}_2, \ldots, \mathbf{x}_\mathcal{T} )$, the goal is to find a reranking function $R$, whose ordering $r(\mathbf{x}_i)_{i=1:\mathcal{T}}$ approximates the optimum ordering $r^*= (d_1, \ldots, d_\mathcal{T})$.
There are multiple ways to rank a list of times, pair-wise ranking is one of them. 
Given a set of ordered pairs of images $O_o= \lbrace (d_i, d_j) \rbrace $, where $d_i > d_j$ i.e. image $i$ has stronger preference to be ranked higher than $j$, and a set of un-ordered pairs of image $O_s= \lbrace (d_i, d_j) \rbrace $, where $d_i \sim d_j$, meaning there is no obvious preference for each one of them. The learned ranking function $R$ should satisfy $R(\mathbf{x}_i) > R(\mathbf{x}_j)$ for all pairs of $\lbrace (\mathbf{x}_i, \mathbf{x}_j): d_i>d_j \rbrace$ for the training set, and also generalize beyond to new dataset.

Given a set of $\mathcal{T}$ items that ranked top ${\mathbf{x}_1, \mathbf{x}_2, \ldots, \mathbf{x}_\mathcal{T}} $, which are highly relevant. The reranking function is defined as:
\begin{align}
\widehat{R}(\mathbf{x}_t)   &= \sum_{i=1}^{\varrho} \mathcal{W}_i \Gamma_i( S_i(\mathbf{x}_t) ) 
\end{align}
\noindent where $S_i(\mathbf{x}_t)$ is the classification function defined in Eq.~\ref{eq:Fc}, which represents the score of  attractiveness. $\mathcal{W}_i$ is the weights learned from the PMF-user choice model in section~\ref{sec:weighing}, which is the level of preference from user. $\varrho$ is number of the top types from the ranked set of ${1, \ldots, i, j, \ldots, \varrho}$, where $\mathcal{W}_{i}>\mathcal{W}_{j}$, and $\varrho \leq c = \| \mathbb{C} \|$. 
We propose three different ways to define $\Gamma$ function, we refer them as PMFP, PMFS and PMFL respectively. 
\allowdisplaybreaks
\begin{align}
\Gamma_i^{PMFP} \left( \mathcal{S}_i(\mathbf{x}_t) \right) &=  \frac{1}{1+e^{\mathcal{S}_j(\mathbf{x}_t)}}   \label{eq:PMFP} \\
\Gamma_i^{PMFS} \left( \mathcal{S}_i(\mathbf{x}_t) \right) &=  \mathcal{S}_j(\mathbf{x}_t) \label{eq:PMFS} \\
\Gamma_i^{PMFL} \left( \mathcal{S}_i(\mathbf{x}_t) \right) &=  \begin{cases}1, & \text{if}~ i=\argmax_{j=1,\ldots, c}S_j(\mathbf{x}_t)  \\ 0, & \text{otherwise} \end{cases} \label{eq:PMFL}
\end{align}

\subsection{Experiment Results \& Evaluations}

The goal of the experiment is to show the benefit of click-through by considering visual element on top of relevance. 
Therefore, to generate evaluation dataset without introducing biases from the variations of the proportion of each display types, we constraint to uniform proir distributions of each display type. We focus only on top 9 items, by randomly selecting 3 images from each types. 

We compare the proposed unsupervised re-ranking schema with two baseline approaches:
1) Assume the order of the top $\mathcal{T}$ highly relevant items are randomly assigned. 
2) Apply supervised learning algorithm, in particular, rankSVM~\cite{joachims2002optimizing} to predict the re-ranking order. 

The Ranking SVM algorithm was originally proposed for search engine optimization for document retrieval. Beginning with the SVM approach, the Ranking SVM uses a method for learning the retrieval function by optimizing a set
of inequalities. By modeling $R$ be a linear function of $\vec{\omega}$ over the feature vector represented by $\mathbf{x}$, the learning becomes computing the weight vector $\vec{\omega}$:
\begin{align}
\text{minimize}~ & L_p(\vec{\omega}, \xi) = \frac{1}{2}  {\parallel \vec{\omega} \parallel }^2 + C \sum \xi_{ij}  \\
\text{s.t.:} ~   &      \vec{\omega} (\mathbf{x}_i, \mathbf{x}_j) \geq 1 - \mathbf{x}_{ij}, \forall \lbrace (\mathbf{x}_i, \mathbf{x}_j): d_i > d_j \rbrace
\end{align}

To train rankSVM, we utilize the scores of attractiveness generated from the trained PMF classifiers (Eq.~\ref{eq:Fc}). Multiple feature combination are experimented. We found the best performance is obtained by using feature vector that contains both label information and classifier decision scores. So the final feature vector representation of each sample is  $\lbrace \mathcal{L}, \mathcal{S}_{j\in (p,m,f)}(\mathbf{x}), g(\mathcal{S}_{j\in (p,m,f)}) \rbrace$, where $g(x)=\frac{1}{1+e^{(x)}}$. We randomly split the data (in total $N$=484 retrieval sessions) by using 75\% for training and 25\% for evaluation. One should note that the training complexity of ranking SVMs is inherently more expensive by two asymptotic orders of magnitude (with respect to data size).

\textit{Measures of performance}. We report the normalized Discounted Cumulative Gain (\textit{nDCG}).
In this reranking case, we focus on \textit{nDCG} at each top position among the re-ranked images.
\textit{nDCG@K} was computed as:
\begin{align}
nDCG\text{@}K &= \frac{DCG\text{@}K}{IDCG\text{@}K}  \\
DCG\text{@}K &= \sum_{i=1}^{K}\frac{2^{rel_i}-1}{log_2(1+i)}
\end{align}
\noindent where $rel_i$ is the relevance level of the item at rank position $i$ and \textit{nDCG@K} is the \textit{DCG@K} for a perfect ranking. In our case, we'll be using a two-point scale for relevance assessment, i.e. $rel_i=y_i$, which takes binary values capturing user click action. We report \textit{nDCG@K} averaged over all experimented reranking groups. 

\begin{figure}[h]
\centering
\begin{tabular}{c}
\epsfig{file=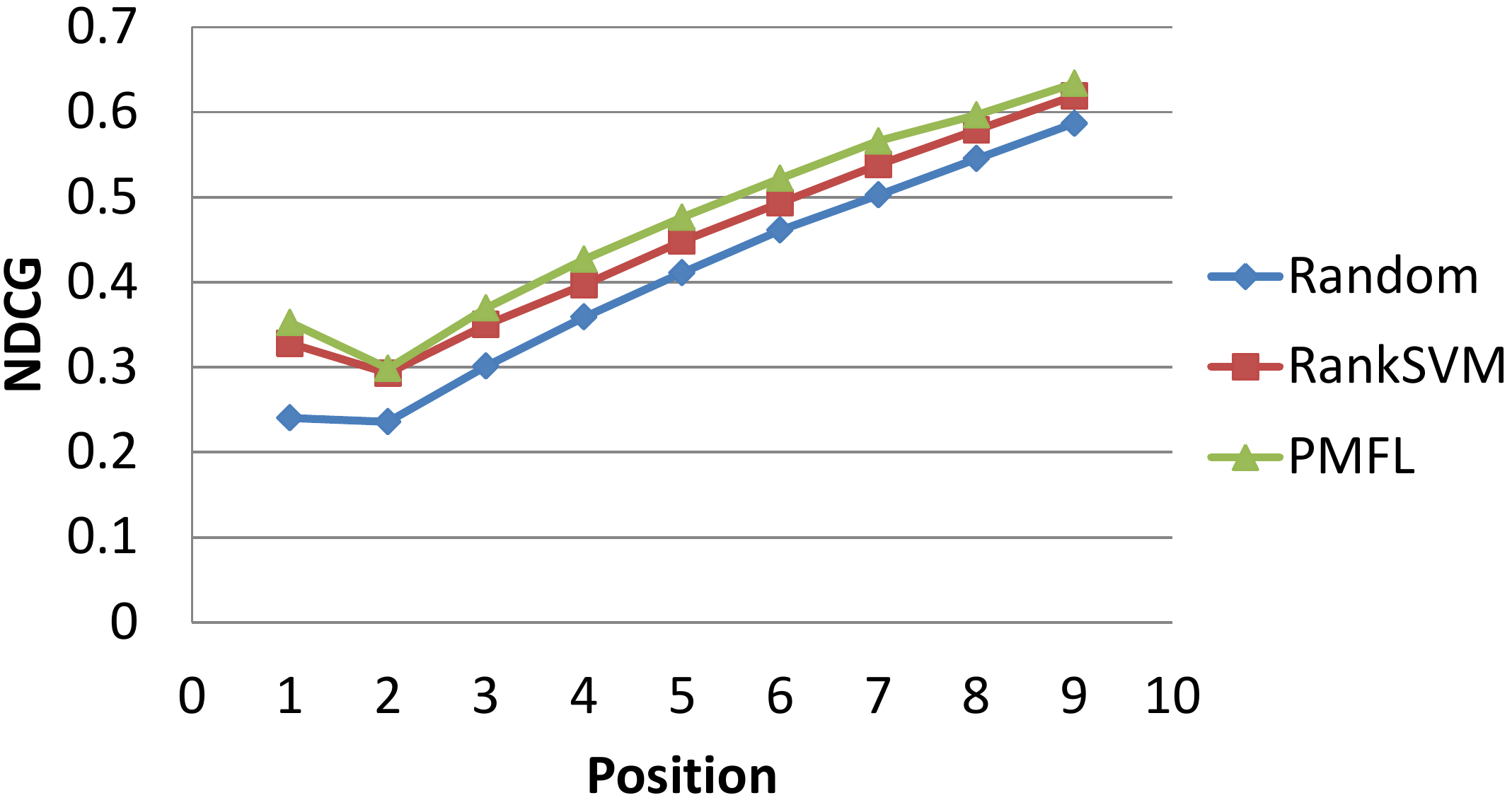,width=0.9\linewidth}  \\
{\small (a) Comparison of PMFL with rankSVM and random assignment.} \\
\epsfig{file=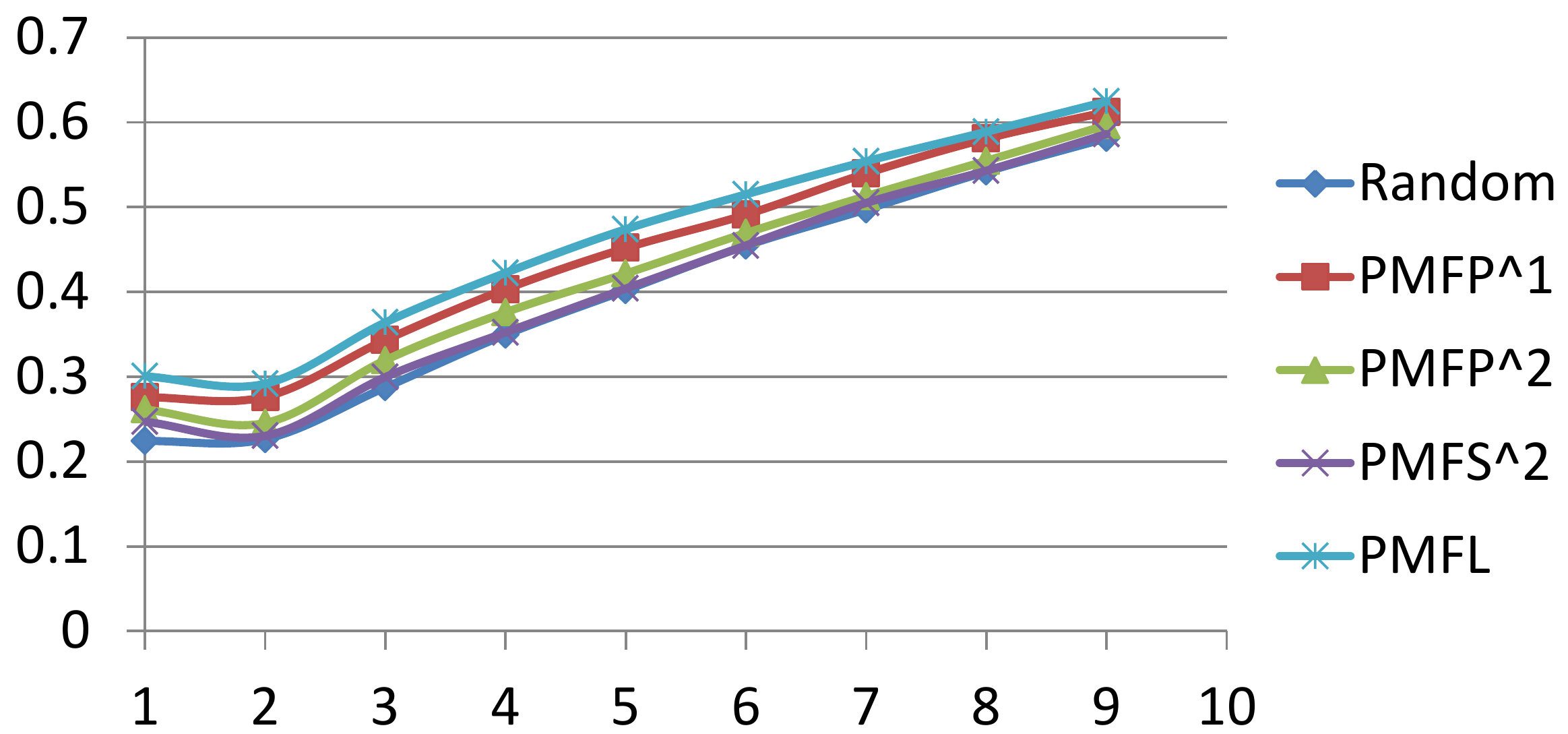,width=.9\linewidth}   \\
{\small (b) Comparison of different PMFx approaches: PMFP, PMFS, PMFL.}  \\
\end{tabular}
\caption{Reranking performance evaluated by \textit{NDCG} and comparison against two baseline approaches: rankSVM and random assignment. Because propose re-ranking schema is unsupervised while rankSVM is supervised, for fair comparison (a) is on testing dataset. (b) is on all available data.}   
\label{fig:NDCG}
\end{figure}

Figure~\ref{fig:NDCG} shows the \textit{nDCG} performance by the proposed algorithm as compared to baseline rankSVM and random assignment. 
Performance of random assignment are computed by averaging more than 10 times of random positioning.
In contrast to rankSVM, which is a supervised algorithm, our propose re-ranking approach is unsurpervised. For fair comparison, in Figure~\ref{fig:NDCG}~(a), we use the performance computed from the same testing dataset to compare with rankSVM. Performance of random assignment is also based on this subset.
In Figure~\ref{fig:NDCG}~(b) we show performances obtained from different design of $\Gamma$ function.
PMFP$~\hat{}~1$ refers to method proposed in Eq.~\ref{eq:PMFP} with $\varrho=1$
Similarly, PMFS$~\hat{}~2$ refers to method proposed in Eq.~\ref{eq:PMFS} with $\varrho=2$ and PMFL refers to method proposed in Eq.~\ref{eq:PMFL} with $\varrho=3$. Since the proposed approach is unsurpervised, the evaluation is done using all the available data. 

\section{Conclusions}
\label{sec:conclusion}

In this paper, we address the effectiveness of product presentation for online fashion market. We conduct two phase study by analyzing both large-scale behavior data (click, watch, purchase) from a real e-commerce market and the user survey data from a more controlled and targeted experiment.
We categorize attractiveness of clothing presentation to a three-point scale, using common display types in fashion field: {\fontsize{8.0}{9.8}\selectfont \textsf{Person}}, {\fontsize{8.0}{9.8}\selectfont \textsf{Mannequin}}, and {\fontsize{8.0}{9.8}\selectfont \textsf{Flat}}. 
User preference is analyzed on this three-scale attractiveness jointly with other selling dimensions such as price and seller type. We also investigate detailed reasons behind user's choice to understand the motives of their decisions. 

Results suggest that attractiveness revealed by using human modeling is the most effective product presentation among the three. Real online market data shows that effective presentation can help to attract user's attention and raise sell-through. Such preference has significant impact on user's decision even when compared to brand or price which are often thought to be highly important factors for clothing category.

In addition, we propose a PMF-user choice model and quantitatively measure user's preference on each of them. A new ranking function, which aims to promoting attractiveness for fashion clothing category is developed in order to improve user engagement on top of relevance. It incorporates both the learned user preferences and presentation efficacy measured by the three-scale of attractiveness. 

This work provides useful insights for apparel e-retailers to design better selling strategy. It also has wide applications such as feeds-like recommendation, advertising. Similar strategy and learning framework is generalizable to other categories and domains where visual presentation is crucial for the product.

\newpage

\bibliographystyle{abbrv}
\bibliography{www-lbd-bib}

\end{document}